\documentclass[a4paper,oneside,onecolumn,british,twocolumn,latin9]{aa}
\usepackage[T1]{fontenc}
\usepackage{babel}
\usepackage{cprotect}
\usepackage{url}
\usepackage{amsmath}
\usepackage{graphicx}
\usepackage[pdfusetitle,
 bookmarks=true,bookmarksnumbered=false,bookmarksopen=false,
 breaklinks=false,pdfborder={0 0 0},pdfborderstyle={},backref=false,colorlinks=false]
 {hyperref}
\hypersetup{
 unicode=true}

\makeatletter

\pdfpageheight\paperheight
\pdfpagewidth\paperwidth

\titlerunning{A space of parameter spaces in the space sciences}
\usepackage{xcolor}

\makeatother

\begin{document}
\title{A space of inference spaces in the space sciences}
\subtitle{Parametric Bayesian inference in astronomy, cosmology and particle
physics}
\abstract{A sample of parametric Bayesian inference applications from astronomy,
cosmology and particle physics is studied, augmented by mock data
sets and toy problems. The parameter spaces of these parametric physical
models and their posterior distributions from analysing specific data
are characterised by (1) the number of model parameters, (2) whether
the posterior shape is similar to a Gaussian, (3) whether the posterior
has light or heavy tails, (4) how small the posterior is compared
to the prior, i.e., how informative the data are, (5) whether some
parameters remain unconstrained while others are highly constrained,
(6) whether the posterior has multiple, disconnected modes, and (7)
whether the inference undergoes phase transitions. These axes define
a parameter space of inference problems. We characterise each of the
inference problems and observe that inference in astrophysics spans
the entire parameter space, from low to high dimensionality, mono-
to multi-modal, and a variety of complex distributions that range
from uninformative to highly informative. Furthermore, the computational
cost of the physical models can range from milliseconds to dozens
of seconds. The collated sample of inference problems is proposed
as a standard test bed for new samplers. For reproducibility and ease
of use, a Docker compute image is provided.}
\author{Johannes Buchner}
\institute{Max Planck Institute for Extraterrestrial Physics, Giessenbachstrasse,
85748 Garching, Germany}
\maketitle

\section{Introduction}

Fitting parametric models to experimental or observational data is
a central method for estimating physical parameters within a specified
statistical model \citep{Dose2003,vonDerLinden2014}. Bayesian data
analysis combines probability models, prior information, and likelihood
functions to quantify parameter uncertainty and compare competing
models \citep{Loredo2024}. In astronomy and other observational sciences,
physical properties are often inferred indirectly from remotely sensed
data. Accurate modelling of the measurement process, including signal
formation, instrumental filtering, noise, calibration uncertainties,
and other nuisance parameters, is therefore essential for obtaining
stable and physically meaningful inferences \citep{Craig1986,Loredo2024,Feigelson2020}.
Some examples include fitting the power spectrum of the Cosmic Microwave
Background (CMB) with Dark Energy and Cold Dark Matter ($\Lambda$CDM)
cosmologies \citep[e.g.,][]{Trotta2008}, fitting time series of the
radial velocity of a host star gravitationally pulled by its exoplanets
\citep[e.g.,][]{Ford2007}, dissecting multiple components in spectra
and population inference from uncertain measurements of many individual
objects, such as luminosity or mass functions. The probable ranges
of model parameters inferred from observational data are often tested
in a Bayesian framework, once prior and likelihood are specified,
by employing Monte Carlo samplers.

Monte Carlo sampling methods of varying complexity have been developed
over several decades. These include importance sampling (IS), Markov
chain Monte Carlo (MCMC), nested sampling and sequential Monte Carlo
(SMC), among other methods \citep{Hesterberg2002,Skilling2004,Bardenet2013}.
Specific implementations of these algorithms can differ in their initialization,
exploration strategy, proposal or transition mechanism, adaptation
scheme, and termination or convergence criteria \citep{Tierney1994,Neal2011,Dai2022,Buchner2021c,Ashton2022}.
The choice and tuning of these components depend on the target distribution
and the computational characteristics of the application, such as
dimensionality, dependence between parameters, multimodality, and
the availability of parallel computation. The reliability of parameter
recovery can be assessed by simulating datasets from a model with
known parameters and then applying the inference procedure to those
datasets \citep[simulation-based calibration;][]{Talts2018,Direnzo2023}.
A complementary strategy is to deliberately construct toy inference
problems that reproduce selected features of the real problem while
being easier to understand and faster to analyse. Because inference
problems differ substantially in their structure, it is useful to
compare different algorithms on the same problem. A preliminary step
towards a holistic understanding of Monte Carlo samplers is to identify
similarities and differences of inference problems. It is the aim
of this work to provide one (not necessarily unique) quantification
of the space of inference problems.

Inference problems differ substantially by the posterior distribution
that a Monte Carlo sampler has to explore. The main characteristics
of problems include (1) the number of model parameters, (2) whether
the posterior shape is similar to a Gaussian, (3) whether the posterior
has light or heavy tails, (4) how small the posterior is compared
to the prior (i.e., how informative the data are), (5) whether some
parameters remain unconstrained while others are highly constrained,
(6) whether the posterior has multiple, disconnected peaks, and (7)
whether the inference undergoes phase transitions. Besides a systematic
classification of inference problems based on the seven characteristics,
this work presents a diverse set of real and toy inference problems
that cover the entire classification space.

\section{Data: collated inference problems}

To cover most of the problem space, we collected inference problems
published in the literature and available in open-source physics packages.
Appendix~\ref{subsec:Real-problems} introduces real-world problems,
which form the main sample in this work. Simplified mock problems
with generated data sets that approximate real-world problems are
presented in Appendix~\ref{sec:Mock-problems}. Appendix~\ref{subsec:Toy-problems}
introduces artificial toy problems. 

For each problem, the likelihood function $L(\theta)$ is defined
together with prior distribution $\pi(\theta)$ over the parameter
space. 

\section{Method: Characterisation of parametric inference}

Various difficulties are encountered by different sub-disciplines.
Here we specify six characteristics and give a mathematical definition
for each. Finally, we present a visual presentation which characterises
the posterior degeneracies.

\subsection{Dimensionality}

In astrophysics, fitting problem dimensionalities range from 1 to
millions parameters. Examples of extremely high-dimensional problems
include non-parametric morphological (image) analyses, and hierarchical
Bayesian models with large number of calibration nuisance parameters.
With more parameters, the possible combinations of parameter values
rise exponentially (the curse of dimensionality). This makes the problem
complex to explore and distances between parameter space points become
less informative. 

Here we define three common sub-groups: low-dimensional ($d=2-9$),
mid-dimensional ($d=10-29$) and high-dimensional ($d\geq30$). The
boundaries are set near where simple and more sophisticated ideas
of geometric sampling start failing. Extremely high dimensions ($d\gg100$)
are not the focus of this work. We note that these virtually always
require the derivatives of likelihood functions to effectively navigate
the parameter space. The availability of likelihood derivatives could
be considered an additional classification category.

\subsection{Information gain (\textquotedblleft Depth\textquotedblright )}

Depending on the data quality of the experiment, the posterior may
be a tiny region of the prior, or be identical with the prior. This
can be quantified by the Kullback-Leibler divergence, or surprise,
between the two probability distributions:

\[
D_{\mathrm{KL}}=\int P(\theta)\log\frac{P(\theta)}{\pi(\theta)}d\theta
\]
Here, $\pi$ and $P$ give the prior and posterior over the parameter
vector space $\theta$. In the case of a base-e logarithm, the unit
of $D_{\mathrm{KL}}$is nats, and approximately means how many e-foldings
it takes to cut the prior until the posterior is reached. Finding
that small region can be a challenge for sampling algorithms (and
maximum likelihood minimisers).

In practice, the information gain is already computed by nested sampling
algorithms internally for error estimation, and we adopt that method
as a measurement. Also, the information gain is related to the number
of iterations of the nested sampling algorithms needs to zoom in until
the likelihood appears flat.

\begin{figure*}
\begin{centering}
\includegraphics[height=4.5cm]{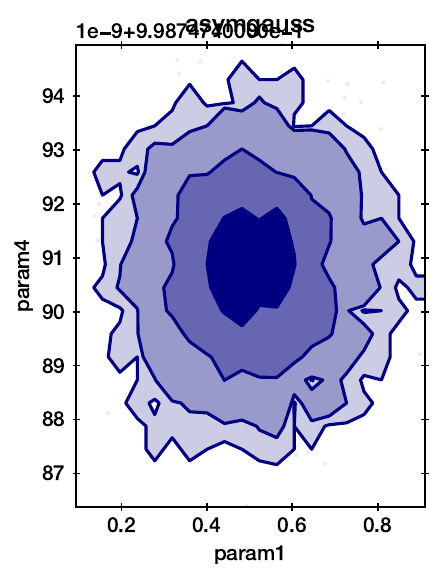}
\includegraphics[height=4.5cm]{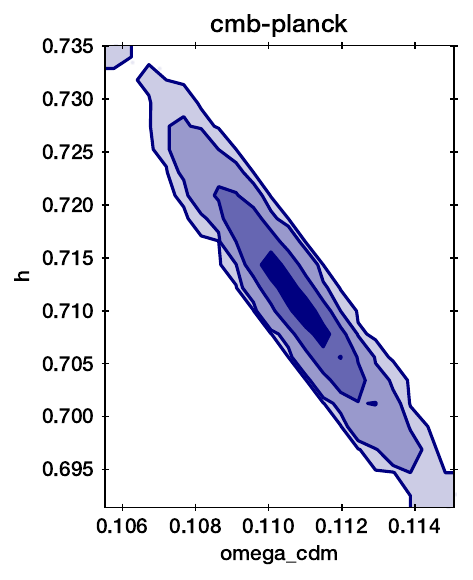}
\includegraphics[height=4.5cm]{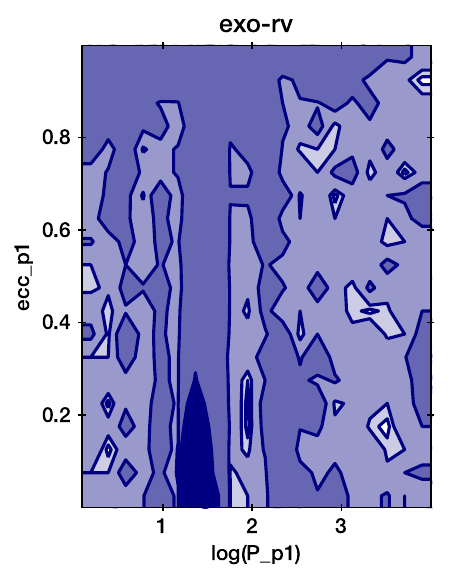}
\includegraphics[height=4.5cm]{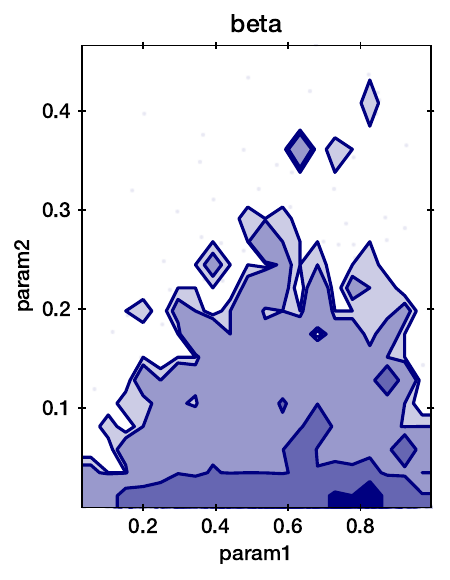}

\includegraphics[height=5cm]{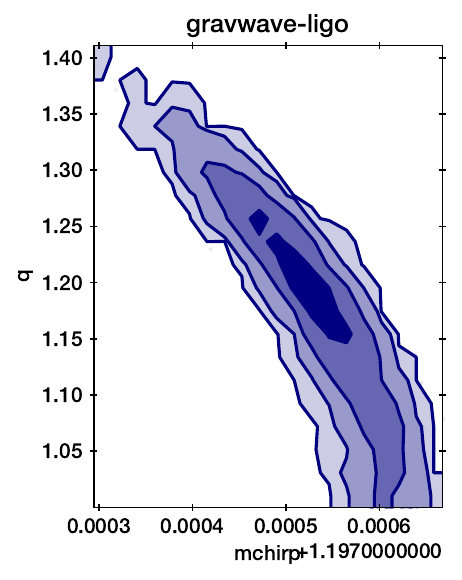}
\includegraphics[height=5cm]{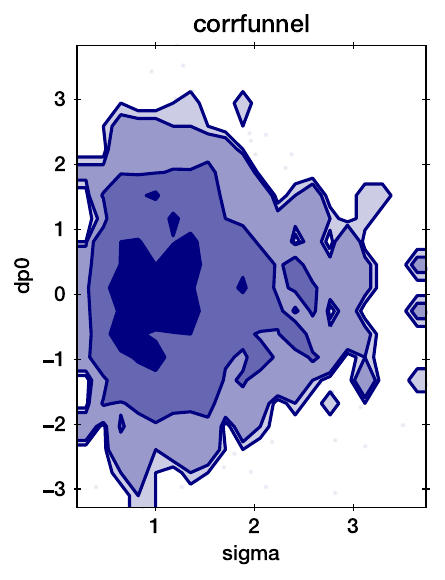}
\includegraphics[height=5cm]{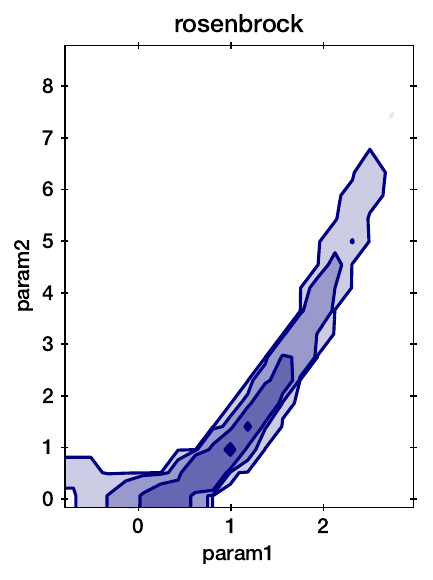}

\includegraphics[height=4.5cm]{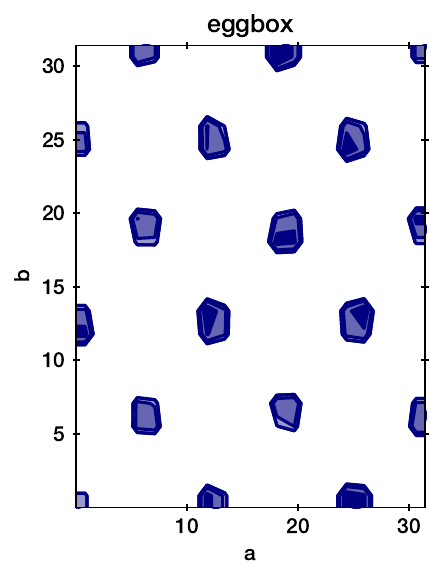}
\includegraphics[height=4.5cm]{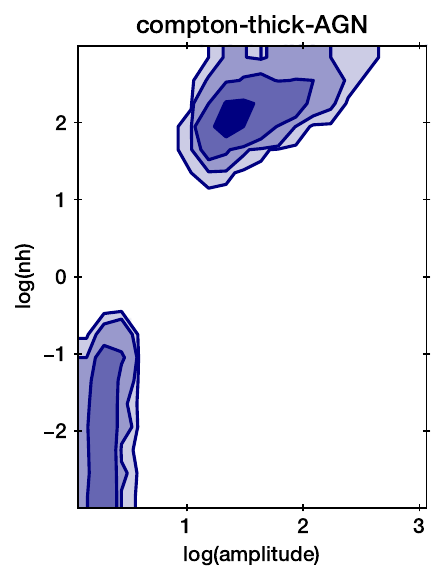}
\includegraphics[height=4.5cm]{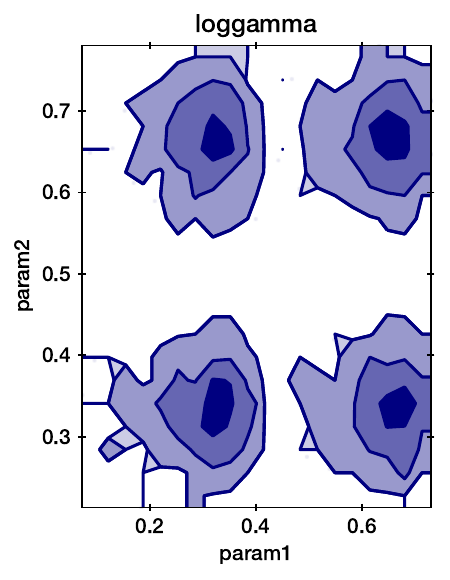}
\includegraphics[height=4.5cm]{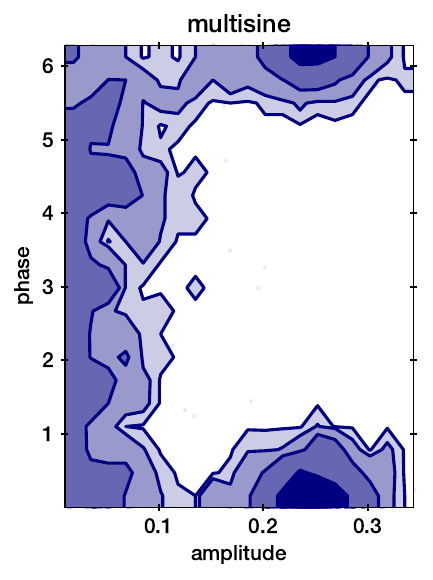}
\par\end{centering}
\caption{\protect\label{fig:pairwise-posterior}Selected pair-wise posterior
distributions from some of our problems. The 1, 2, 3 and 4-sigma equivalent
probability mass contours \citep[made with corner;][]{Foreman-Mackey2016}
illustrate non-linear degeneracies (e.g., middle panels), unequal
axes (e.g., left-most and right-most top panels), multi-modality (bottom
panels). The loggamma problem (third panel, bottom row) also has heavy
tails towards the left. Some parameters are uninformative (e.g., param1
in top right panel) or at the prior parameter edge (middle left panel,
ratio $q\protect\geq1$, top right panel, $0\protect\leq\mathrm{param2}\protect\leq1$).}
\end{figure*}

\subsection{Modes}

When data can be explained with similar quality by different combinations
of processes, the posterior exhibits multiple peaks. This is common
in fits of multiple components with (nearly-)interchangeable predictions,
paired with poor discrimination power of the data. Algorithms based
on local jumps can find it difficult to navigate between modes, because
a proposal tuned to a single mode may reach another distant mode with
vanishing proposal probability. The bottom row of Figure~\ref{fig:pairwise-posterior}
shows examples of multimodal distributions.

In the literature, a vast diversity of definitions exist. Mixtures
of Gaussian components can be fitted to samples, and the needed number
of components determined \citep[e.g.,][]{McLachlan1987,Keribin2000}.
However, the number of mixture components may indicate skewness and
heavytailed distributions rather than the number of modes, which depends
on component overlap and covariance structure \citep{Ray2005}. Given
a smooth fitted density, modality can be defined through local maxima
or basins \citep[e.g.][]{Genovese2016,chacon2019mixture}. Here, however,
we are primarily interested in low-probability regions that isolate
high-probability regions from each other. Such gaps are complex degeneracies
that are physically interesting, but also of practical interest, as
Monte Carlo samplers may find them difficult to traverse. 

To mathematically define multi-modality, a threshold criterion is
needed to define disconnectedness. For simplicity and reproducibility,
we adopt a very simple, heuristic approach. First, histograms of the
marginal posteriors in physical coordinates are histogrammed into
20 bins. Bins with density less than 1/5 of the peak density are considered
``empty''. Gaps are identified, and thresholds that bracket the
peaks extracted. This is repeated for every dimension. Then, all combinations
of brackets are computed. These are the clusters. If posterior samples
are members of multiple clusters, the clusters are merged. The number
of remaining, non-empty clusters is the number of modes of the problem.
A more rigorous (and likely more complex) approach of mode counting
is left for future work.

\subsection{Non-Gaussianity}

\begin{figure*}

\includegraphics[width=1\textwidth]{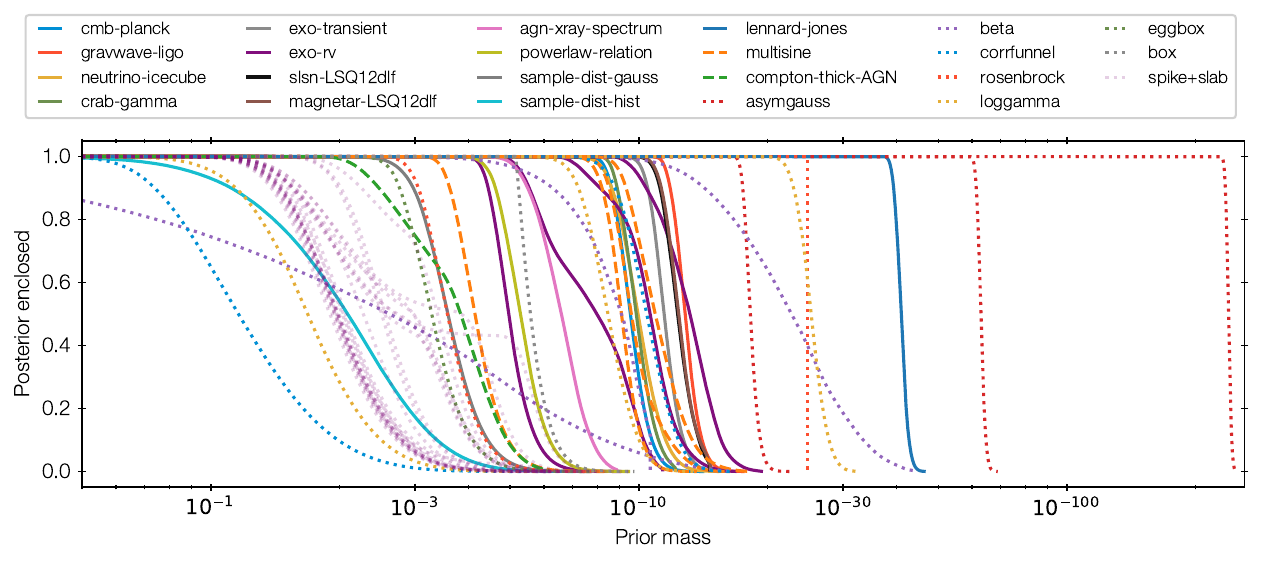}

\caption{\protect\label{fig:volcurve}Probability fraction enclosed as a function
of prior volume. From the highest likelihood regions outwards, the
volume is increased from left to right until the entire prior space
($V=1$) is enclosed. The median (cross) indicates how large the posterior
volume is relative to the prior, and is related to the \emph{information
gain}. Quantiles at 5\% and 95\% indicated as circles indicate the
shell where most probability mass is enclosed. This is related to
the \emph{tail weight}. Some problems show wide transitions (e.g.,
green dashed) relative to a Gaussian (blue dashed).}

\end{figure*}

The Bernstein--von Mises theorem states that the posterior is shaped
like a multi-variate Gaussian for many classes of models if the number
of data is large. Then, Laplace's approximation of the posterior with
a log-quadratic density function is justified. This can furthermore
occur if the model is linear in its parameters, or a first-order Taylor
expansion of the model provides a reasonable approximation at the
maximum a posteriori. For this reason, some algorithms are constructed
to behave optimally when the posterior is Gaussian (see e.g., Laplace
approximation).

The Gaussianity of a posterior can be easily measured through the
surprise from a best-fitting multi-variate Gaussian $G=N(\mu,\Sigma)$
to the posterior, $D_{KL}(P||G)=\int P(\theta)\log\frac{P(\theta)}{\pi(\theta)}d\theta$.
We obtain a suitable Gaussian from the mean $\mu$ and covariance
$\Sigma$ of the posterior samples. The result is dependent on the
parameterisation. We use posterior samples in the unit cube, so that
when the prior dominates, regardless of the chosen prior distribution,
the posterior is not Gaussian but uniformly distributed. We then define
the non-Gaussianity, normalised by the number of parameters, as: 
\[
NG=D_{KL}(P||G)\thinspace/\thinspace d
\]
In multi-modal cases, the posterior is highly non-Gaussian, $G$ is
a poor approximation, and thus $NG$ is high. 

\subsection{Tail weight (\textquotedblleft Width\textquotedblright )}

\begin{figure}

\includegraphics[width=1\columnwidth]{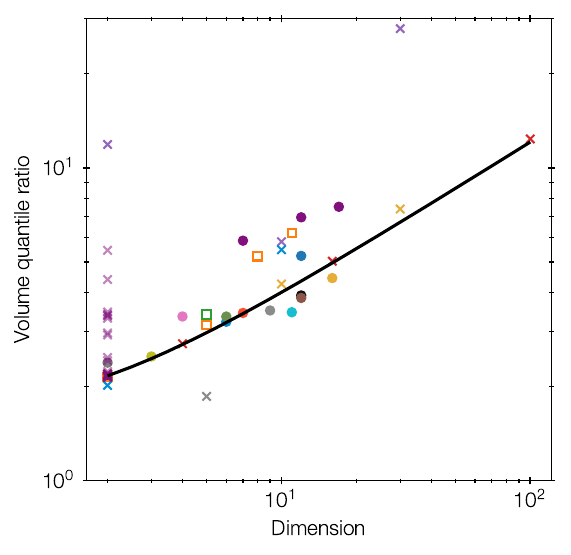}

\caption{\protect\label{fig:width}The ratio of $97.5\%$ and $2.5\%$ quantiles
of the volume compared to dimensionality. Red crosses are Gaussian
likelihoods, which follow the quantile ratio of a $\chi_{d}^{2}$-distribution
(black curve). We use the black curve to normalise.}
\end{figure}

While the posterior may have ellipsoidal contours like a Gaussian,
the posterior density may decline steeper or shallower than a square-exponential,
i.e., have thin or heavy tails. For example, when outliers are allowed
(e.g., in Student-t distribution or explicit outlier modelling), the
wings of the posterior can be wide. Some algorithms may be optimized
for square-exponential declines. Another cause of heavy tails is when
most data points are fitted well by one component, and a minor component
relevant for a small data subset improves the fit slightly (for example,
in blind spectral line searches on top of a continuum). This leads
to a phase transition, where the parameter space to be explored changes
rapidly (with small likelihood change) from a wide volume to a narrow
volume. This is difficult for many samplers.

To quantify how the weight is distributed over the prior volume, we
compute the prior volume above a given likelihood threshold. This
is illustrated in Figure~\ref{fig:volcurve}. We compute the $2.5\%$
and $97.5\%$ quantiles of the volume ranges where most of the probability
mass resides. Figure\,\ref{fig:width} shows $V_{97.5\%}/V_{2.5\%}$
for our inference problems. The expectation from Gaussian posterior
shells, from quantiles of the chi-square distribution with d degrees
of freedom, is:

\[
\mathrm{TW}_{0}=\frac{F_{\chi_{d}^{2}}^{-1}(0.975)}{F_{\chi_{d}^{2}}^{-1}(0.025)}\approx1.2\times\sqrt{d}
\]
Normalising by this, we compute the tail weight as:
\[
\mathrm{TW}=\log_{10}\left(\frac{V_{97.5\%}}{V_{2.5\%}}\right)/\mathrm{TW}_{0}
\]

In practice, the mapping of volume and likelihood, as well as the
normalising constant, the marginal likelihood integrated over volume
shrinkages, are already computed by nested sampling. We cap the value
at $\mathrm{TW}=7$.

\subsection{Inequality}

Some model parameters may alter the model prediction strongly, while
others have more minute implications. Because of this, the posterior
of some parameters can be consistent with the prior with no information
learnt. For other parameters, its plausible range may have diminished
by several orders of magnitude. Proposals that are isotropic over
the parameter space directions then may perform poorly. The top row
of Figure~\ref{fig:pairwise-posterior} shows such examples in pairs
of parameters. 

We quantify the inequality of parameters in problems, as:

\[
\mathrm{IE}=\frac{\max_{i}\left\{ \max\left(\mathrm{1\,bit},\mathrm{IG}_{i}\right)\right\} }{\min_{i}\left\{ \max\left(\mathrm{1\,bit},\mathrm{IG}_{i}\right)\right\} }.
\]
Here, the information gain $\mathrm{IG}_{i}$ in bits for each parameter
$\theta_{i}$ is computed from the prior $\pi(\theta_{i})$ and posterior
$P(\theta_{i}|D)$ marginal distributions as: 
\[
\mathrm{IG}_{i}=\int P(\theta_{i}|D)\log_{2}\frac{P(\theta_{i}|D)}{\pi(\theta_{i})}d\theta_{i}
\]
If all parameters have similar posterior uncertainties, $\mathrm{IE}\approx1$.
We cap this value to at most $\mathrm{IE}=100$. 

\subsection{Phase transitions}

\begin{figure}

\includegraphics[width=1\columnwidth]{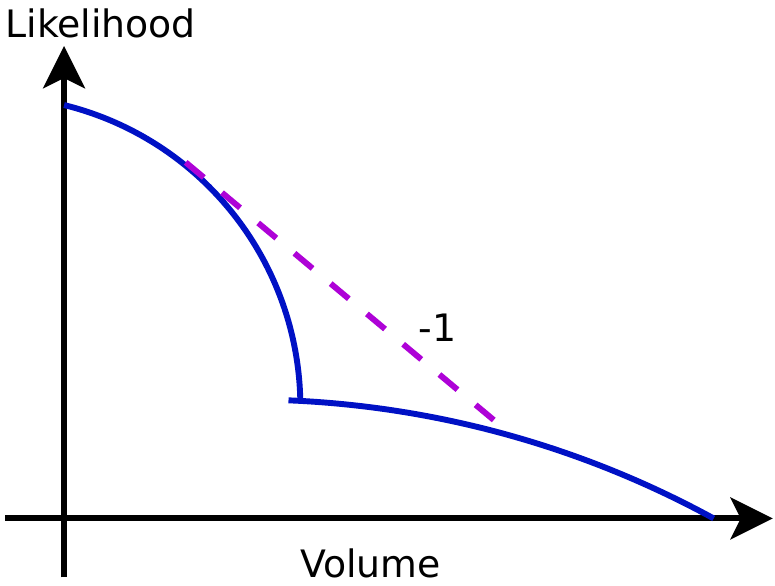}

\caption{\protect\label{fig:phasetransition}In the prior volume - likelihood
plot of nested sampling, phase transitions are visible as non-convexities.
The dashed curve illustrates our technique for closing the curve. }

\end{figure}

Phase transitions are sudden, unexpected changes in volume when the
likelihood is changed continuously. An example is shown in Figure~\ref{fig:phasetransition}.
This is common in multi-component fitting, where the rough overall
data is first approximated, and then fine tuning a second component
onto a subset of the data leads to a rapid further increase in the
likelihood. The situation has been likened in \citet{skilling2009nested}
to condensation of a water vapor into liquid water, where small temperature
changes leads to a sudden change in volume. In the thermodynamics
analogy of Monte Carlo inference algorithms such as simulated annealing
and nested sampling, \citet{skilling2009nested} discussed that in
phase transitions the slope of the likelihood-volume curve deviates
strongly from a power law with index -1. This can affect some inference
techniques, because the relevant geometry of the (proposal) space
before and after the transition is radically different. 

Whether a phase transition is important and even experienced depends
on the Monte Carlo algorithm. Nested sampling needs to navigate phase
transitions across the entire range of likelihood $L$ values, while
Markov Chain Monte Carlo algorithms may primarily navigate the posterior
bulk and thus be insensitive to low-$L$ phase transitions. Here,
our aim is to focus on phase transitions that impact the bulk posterior.

We measure the presence and the impact of phase transitions. Figure~\ref{fig:phasetransition}
illustrates a phase transition in blue, with a wide plateau followed
by a spike of small volume. Here, from each point on the curve we
extrapolate a power law with index -1 rightward (purple in Figure~\ref{fig:phasetransition})
and use it to raise all points to the right of the curve to the extrapolation.
From both the modified curve $L'(V)$ and the original curve $L(V)$,
we compute the evidence, and consider the evidence ratio as our phase
transition indicator:
\[
\Phi=\frac{\int L'(V)\,dV}{\int L(V)\,dV}
\]
Here the denominator with $L$ is the normal nested sampling computation
of the evidence, while the numerator contains the modified curve,
$L'$, as described above. This therefore measures the impact of the
phase transition on the posterior. By construction, $\Phi\geq1$.
If no phase transitions are present, $\Phi\approx1$, while $\Phi>1$
indicates phase transitions.

\subsection{Compact visualisation of posterior structure}

\label{subsec:Compact-visualisation-of}

\begin{figure*}

\includegraphics[viewport=0bp 0bp 1120bp 1136bp,width=1\textwidth]{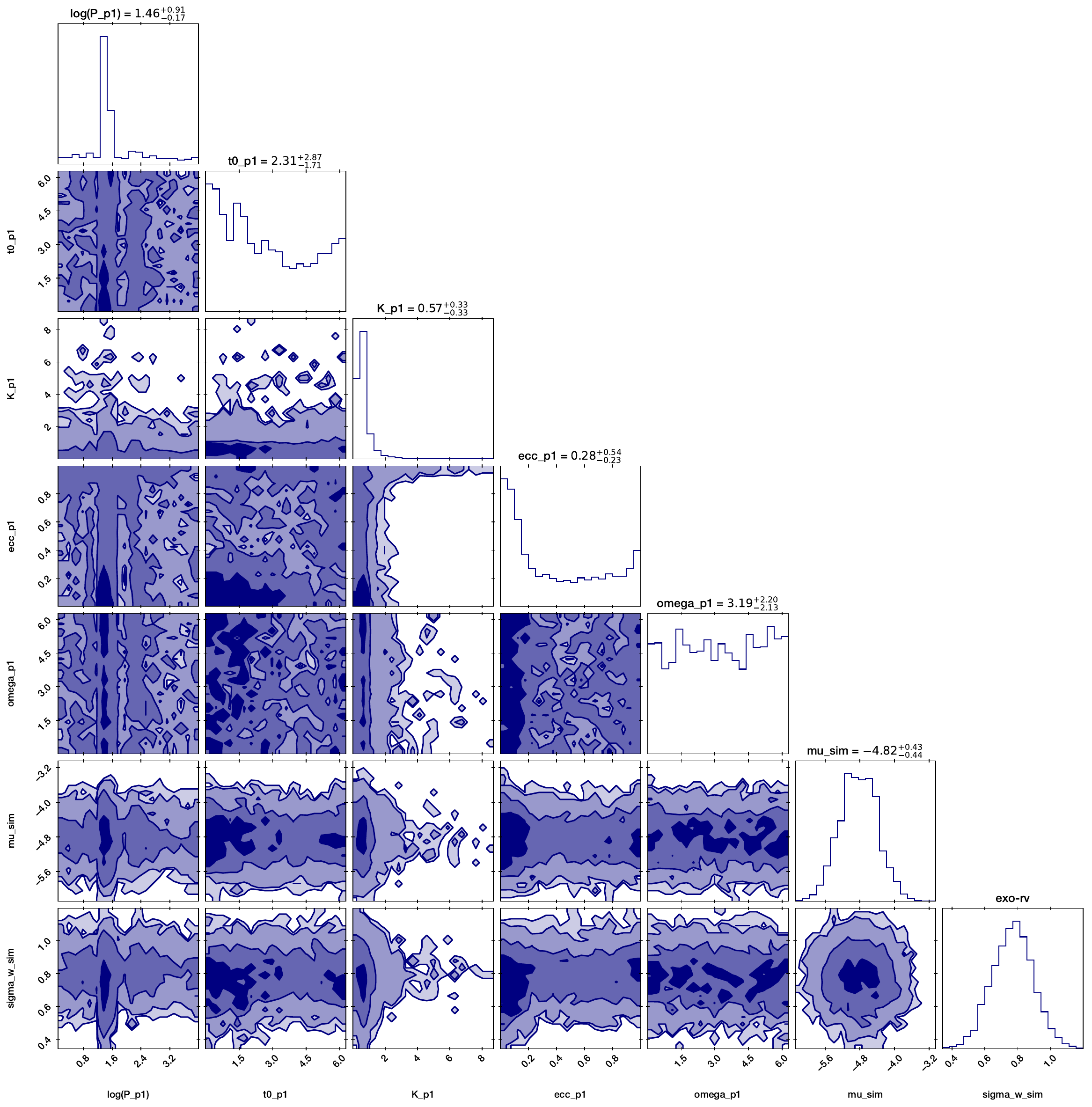}

\caption{\protect\label{fig:corner-example-1}Corner plot \citep[created with corner.py;][]{Foreman-Mackey2016}
for the exo-rv problem with 1 planet. The degeneracies can be non-linear
(see e.g., ecc\_p1 and K-p1). Some parameters are uninformative (omega\_p1),
while others are very well constrained.}
\end{figure*}

For a visual impression of the experienced parameter space and its
linear and non-linear degeneracies, a common tool are corner plots.
These are also known in the statistics literature as pairs plots,
and an example is shown in Figure~\ref{fig:corner-example-1}. Here,
we quantify the further interaction between pairs of parameters and
present a compact visualisation as a $d\times d$ matrix. The diagonal
entries are filled with the information gain $H$ expressed in bits.
The upper and lower triangles are populated with correlation coefficients
of pairs of parameters.

Corner plots only consider the posterior distribution. However, to
express the experience of a sampler fairly, the characterisation should
consider also the prior. Here we work with the posterior samples expressed
with coordinates on the prior cumulative probability distribution
(probability integral transform). In the case of nested samplers working
with prior transforms based on the unit hyper-cube, this merely means
we take the un-transformed posterior samples. First, we characterise
the linear degeneracies. the pair-wise Pearson correlation coefficient
$\rho_{\mathrm{Pearson}}$ is calculated for each posterior parameter
pair. Secondly, we characterise residual non-linear degeneracies.
On each parameter pair, we compute the Spearman correlation coefficient
$\rho_{\mathrm{Spearman}}$ of ordinary least square residuals. Since
the least square regression has direction, for each pair we adopt
the lower $\rho_{\mathrm{Spearman}}$ of the two possible directions.

\begin{figure}
\begin{centering}

\includegraphics[viewport=0bp 0bp 294bp 297bp,width=0.8\columnwidth]{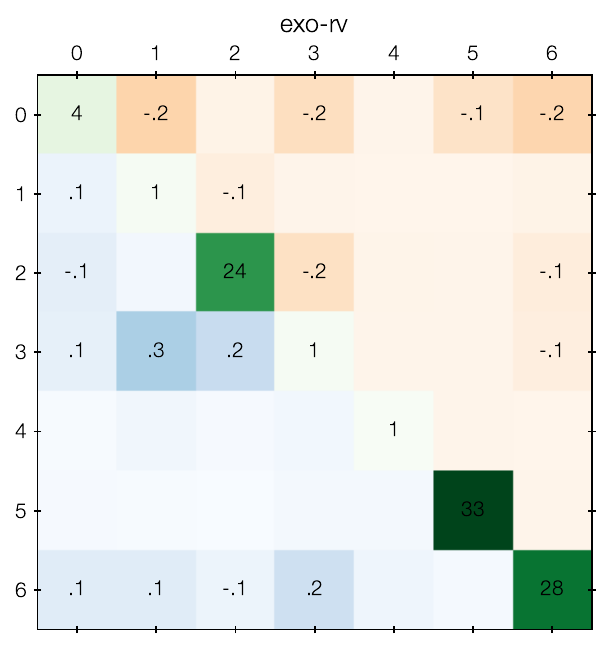}
\par\end{centering}
\caption{\protect\label{fig:degeneracy-viz-example}Proposed visualisation
of the degeneracies, corresponding to Figure~\ref{fig:corner-example-1}.
The lower left triangle in blue shows linear correlation between parameter
posteriors ($|\rho_{\mathrm{Pearson}}|$ from 0 to 1), with darker
colour indicating stronger correlation. Similarly, the upper right
triangle in orange shows Spearman rank correlation between parameter
posteriors, after removing the linear correlations. The diagonal entries
in green indicate the information gain on the parameter, increasing
from white to darker colours. Here, some parameters are uninformative
while others are highly informative.}

\begin{centering}

\includegraphics[viewport=0bp 0bp 299bp 297bp,width=0.8\columnwidth]{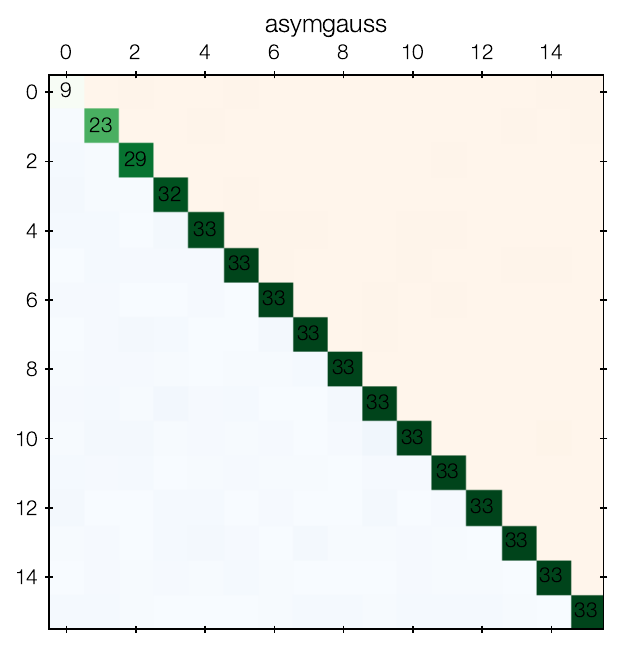}
\par\end{centering}
\caption{Same as Figure~\ref{fig:degeneracy-viz-example}, but for our 16
dimensional Gaussian toy problem. Here, the non-diagonal entries are
near zero, indicating (correctly) that there are no parameter degeneracies.
The diagonal elements vary, with the first parameter being much less
informative than the others.}

\end{figure}

The obtained matrix visualisation is demonstrated in Figure~\ref{fig:degeneracy-viz-example},
corresponding to the corner plot in Figure~\ref{fig:corner-example-1}.
The diagonal shows the parameter information gain, the lower left
entries indicates the linear degeneracies and the upper right entries
indicates the non-linear degeneracies. In the case of a uncorrelated
Gaussian, illustrated in the right panel of Figure~\ref{fig:degeneracy-viz-example},
the off-diagonal elements are zero.

\begin{table*}
\begin{centering}

\begin{tabular}{ll|rrcccccc}
field                & name                 & dim & cost & depth & width & modes & asym & !gauss & phase \\
\hline
\hline
cosmology            & cmb-planck           & 6   & 659  & 1.59 & 1.1 &    1 & 1.34 & 0.01 & 0.003 \\
gravitational waves  & gravwave-ligo        & 7   & 4    & 1.83 & 1.1 &    1 & 2.14 & 0.03 & 0.004 \\
astroparticle        & neutrino-icecube     & 16  & 146  & 0.62 & 0.9 &    1 & 29.68 & 0.12 & 0.004 \\
supernova remnants   & crab-gamma           & 6   & 76   & 1.64 & 1.1 &    1 & 3.27 & 0.06 & 0.006 \\
exoplanets           & exo-transient        & 9   & 4    & 1.27 & 1.0 &    1 & 33.09 & 0.07 & 0.003 \\
                     & exo-rv-0             & 2   & 3    & 2.48 & 1.3 &    1 & 1.19 & 0.01 & 0.000 \\
                     & exo-rv-1             & 7   & 3    & 1.14 & 1.8 &    2 & 33.09 & 0.42 & 0.225 \\
                     & exo-rv-2             & 12  & 2    & 0.88 & 1.7 &    3 & 33.07 & 0.32 & 0.015 \\
                     & exo-rv-3             & 17  & 3    & 0.77 & 1.5 &    9 & 6.16 & 0.13 & 0.031 \\
transients           & slsn-LSQ12dlf        & 12  & 16   & 1.03 & 0.9 &    8 & 14.22 & 0.15 & 0.023 \\
                     & magnetar-LSQ12dlf    & 12  & 14   & 1.03 & 0.9 &    7 & 17.31 & 0.17 & 0.020 \\
extragalactic        & agn-xray-spectrum    & 4   & 4    & 1.65 & 1.4 &    1 & 3.86 & 0.19 & 0.007 \\
                     & powerlaw-relation    & 3   & 3    & 1.76 & 1.2 &    1 & 1.18 & 0.01 & 0.001 \\
                     & sample-dist-gauss    & 2   & 1    & 1.79 & 1.4 &    1 & 1.03 & 0.13 & 0.001 \\
                     & sample-dist-hist     & 11  & 3    & 0.19 & 0.9 &    1 & 11.51 & 0.13 & 0.005 \\
materials            & lennard-jones-6      & 12  & 0    & 3.42 & 1.3 &    1 & 1.00 & 0.11 & 0.003 \\
\hline
mock                 & multisine-0          & 2   & 1    & 2.07 & 1.3 &    1 & 1.51 & 0.01 & 0.000 \\
                     & multisine-1          & 5   & 0    & 1.82 & 1.2 &    2 & 2.00 & 0.03 & 0.002 \\
                     & multisine-2          & 8   & 0    & 1.20 & 1.5 &   20 & 32.43 & 0.50 & 0.002 \\
                     & multisine-3          & 11  & 0    & 0.99 & 1.6 &   67 & 33.04 & 0.54 & 0.006 \\
                     & compton-thick-AGN    & 5   & 1    & 0.77 & 1.3 &    2 & 7.42 & 0.05 & 0.037 \\
\hline
toy                  & asymgauss-4d         & 4   & 0    & 4.55 & 1.1 &    1 & 3.07 & 0.01 & 0.001 \\
                     & asymgauss-16d        & 16  & 0    & 3.93 & 1.0 &    1 & 3.79 & 0.09 & 0.002 \\
                     & asymgauss-100d       & 100 & 0    & 2.38 & 1.0 &    1 & 3.01 & 0.31 & 0.012 \\
                     & beta-2d              & 2   & 0    & 1.31 & 7.2 &    1 & 11.78 & 12.88 & 0.010 \\
                     & beta-10d             & 10  & 0    & 0.91 & 1.5 &   16 & 4.50 & 0.05 & 0.014 \\
                     & beta-30d             & 30  & 0    & 0.76 & 4.2 & 1024 & 12.19 & 1.04 & 0.039 \\
                     & corrfunnel-2d        & 2   & 0    & 0.58 & 1.2 &    1 & 4.82 & 0.01 & 0.001 \\
                     & corrfunnel-10d       & 10  & 14   & 1.04 & 1.4 &    1 & 1.65 & 0.29 & 0.004 \\
                     & rosenbrock-2d        & 2   & 0    & 1.78 & 1.3 &    1 & 1.16 & 0.00 & 0.001 \\
                     & rosenbrock-20d       & 20  & 0    & 1.24 & 0.0 &    1 & 1.00 & 0.03 & 0.000 \\
                     & loggamma-2d          & 2   & 0    & 0.84 & 1.3 &    4 & 1.15 & 0.01 & 0.001 \\
                     & loggamma-10d         & 10  & 0    & 0.85 & 1.1 &    4 & 1.76 & 0.01 & 0.005 \\
                     & loggamma-30d         & 30  & 6    & 0.84 & 1.1 &    4 & 1.82 & 0.12 & 0.007 \\
                     & eggbox-2d            & 2   & 0    & 1.64 & 1.3 &   18 & 1.00 & 0.01 & 0.001 \\
                     & box-5d               & 5   & 0    & 1.12 & 0.7 &    1 & 1.00 & 0.36 & 0.000 \\
                     & spikeslab-1-2d-4     & 2   & 2    & 1.13 & 1.5 &    1 & 1.02 & 0.20 & 0.001 \\
                     & spikeslab-1-2d-40    & 2   & 2    & 1.48 & 2.0 &    1 & 1.00 & 0.84 & 0.170 \\
                     & spikeslab-1-2d-400   & 2   & 2    & 1.92 & 2.6 &    1 & 1.00 & 1.48 & 0.373 \\
                     & spikeslab-1-2d-4000  & 2   & 2    & 1.35 & 3.2 &    1 & 1.01 & 2.36 & 0.478 \\
                     & spikeslab-40-2d-4    & 2   & 2    & 0.99 & 1.3 &    1 & 1.01 & 0.03 & 0.002 \\
                     & spikeslab-40-2d-40   & 2   & 2    & 0.99 & 1.4 &    1 & 1.00 & 0.34 & 0.001 \\
                     & spikeslab-40-2d-400  & 2   & 2    & 0.97 & 1.7 &    1 & 1.01 & 1.15 & 0.003 \\
                     & spikeslab-40-2d-4000 & 2   & 2    & 1.03 & 1.9 &    1 & 1.01 & 2.25 & 0.016 \\
                     & spikeslab-1000-2d-4  & 2   & 2    & 0.95 & 1.3 &    1 & 1.00 & 0.01 & 0.001 \\
                     & spikeslab-1000-2d-40 & 2   & 2    & 1.02 & 1.3 &    1 & 1.00 & 0.07 & 0.001 \\
                     & spikeslab-1000-2d-400 & 2   & 2    & 0.99 & 1.3 &    1 & 1.01 & 0.16 & 0.001 \\
                     & spikeslab-1000-2d-4000 & 2   & 2    & 1.01 & 1.3 &    1 & 1.00 & 0.80 & 0.001 \\
                     & spikeslab-1-2d-40-off1 & 2   & 2    & 1.49 & 2.0 &    1 & 1.01 & 0.88 & 0.186 \\
                     & spikeslab-1-2d-40-off2 & 2   & 2    & 1.28 & 2.0 &    6 & 1.00 & 1.03 & 0.150 \\
                     & spikeslab-1-2d-40-off4 & 2   & 2    & 1.37 & 2.0 &    2 & 1.02 & 0.90 & 0.195 \\
                     & spikeslab-1-2d-40-off10 & 2   & 2    & 1.69 & 1.7 &    2 & 1.00 & 0.40 & 0.030 \\
                     & spikeslab-40-2d-40-off1 & 2   & 2    & 0.98 & 1.3 &    1 & 1.01 & 0.11 & 0.001 \\
                     & spikeslab-40-2d-40-off2 & 2   & 2    & 0.99 & 1.3 &    1 & 1.00 & 0.01 & 0.002 \\
                     & spikeslab-40-2d-40-off4 & 2   & 2    & 0.99 & 1.3 &    1 & 1.00 & 0.02 & 0.001 \\
                     & spikeslab-40-2d-40-off10 & 2   & 2    & 1.28 & 1.3 &    1 & 1.01 & 0.02 & 0.001 \\
                     & spikeslab-1000-2d-40-off1 & 2   & 7    & 0.97 & 1.3 &    1 & 1.00 & 0.02 & 0.001 \\
                     & spikeslab-1000-2d-40-off2 & 2   & 7    & 0.96 & 1.3 &    1 & 1.03 & 0.00 & 0.001 \\
                     & spikeslab-1000-2d-40-off4 & 2   & 7    & 0.98 & 1.3 &    1 & 1.00 & 0.00 & 0.001 \\
                     & spikeslab-1000-2d-40-off10 & 2   & 7    & 1.25 & 1.3 &    1 & 1.00 & 0.00 & 0.001 \\
\end{tabular}
\par\end{centering}
\caption{\protect\label{tab:List-of-problems}List of problems analysed. The
columns describe (1) research field, (2) name, (3) dimensionality,
(4) model evaluation cost in milliseconds, (5) information gain, (6)
tail weight, (7) parameter inequality, (8) Gaussian approximation
information loss, (9) phase transition.}
\end{table*}

\begin{figure*}

\includegraphics[viewport=0bp 0bp 1045bp 1001bp,width=1\linewidth]{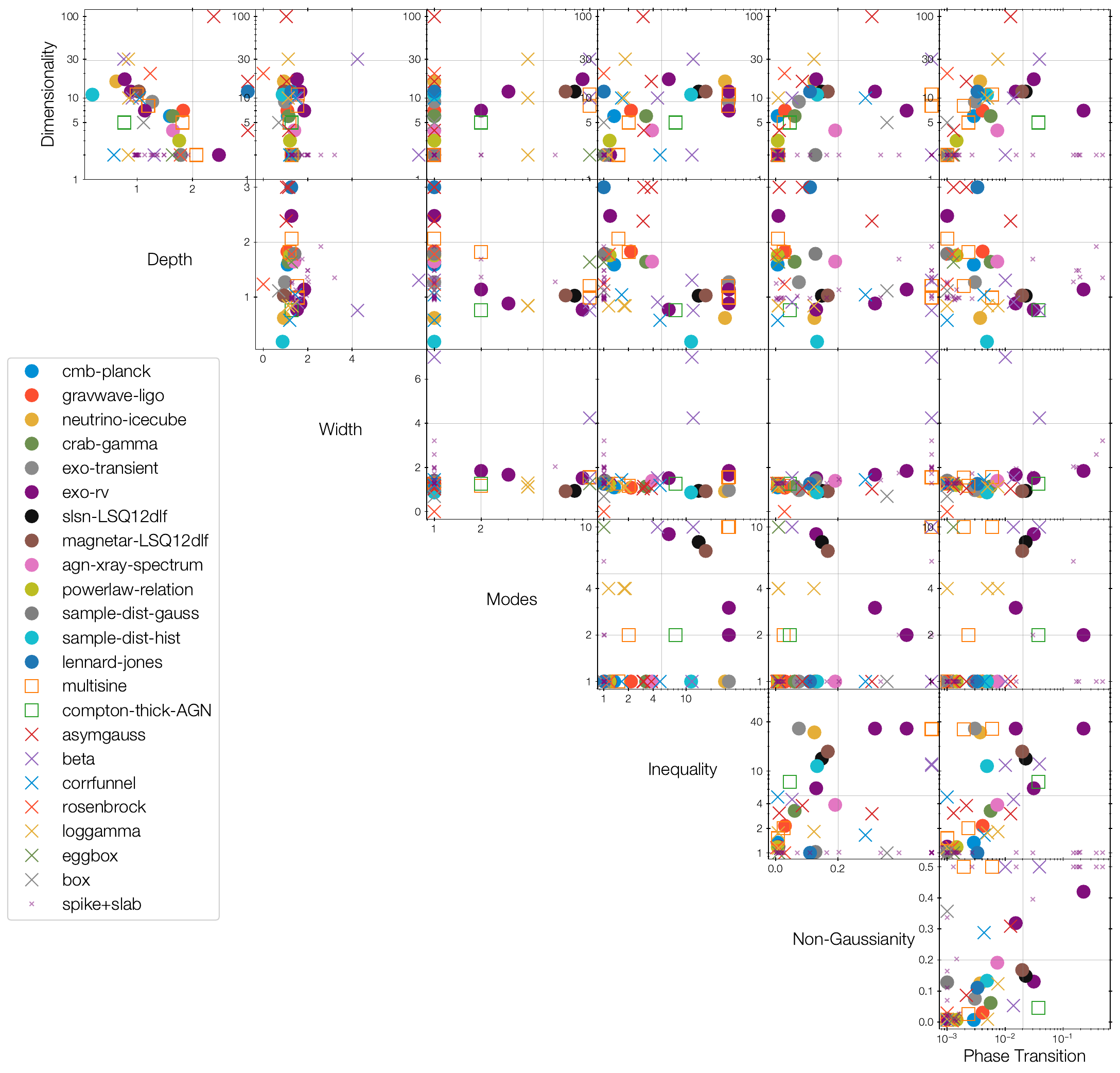}

\caption{\protect\label{fig:Models}Parameter space of the problems over the
defined six characteristics.}
\end{figure*}

\section{Results}

\begin{figure*}

\includegraphics[viewport=0bp 0bp 294bp 297bp,width=0.5\columnwidth]{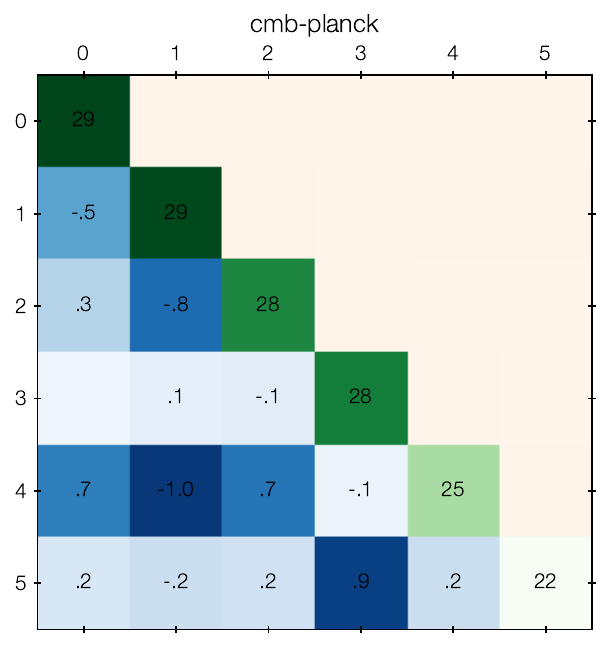}
\includegraphics[viewport=0bp 0bp 294bp 297bp,width=0.5\columnwidth]{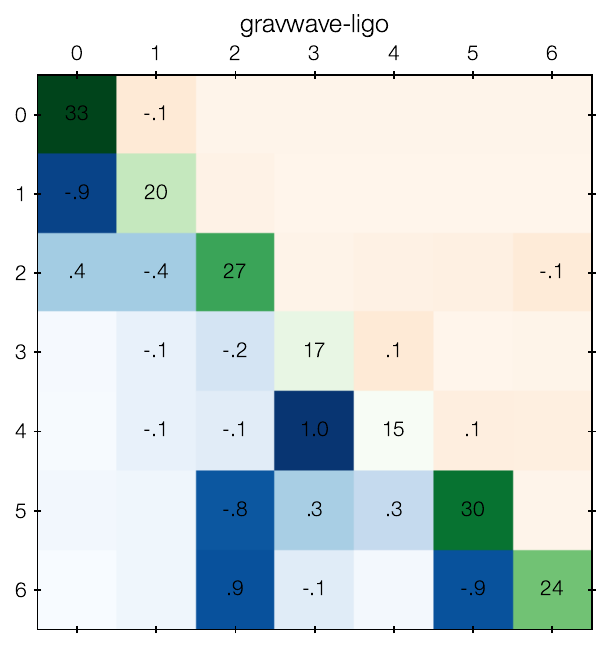}
\includegraphics[viewport=0bp 0bp 294bp 297bp,width=0.5\columnwidth]{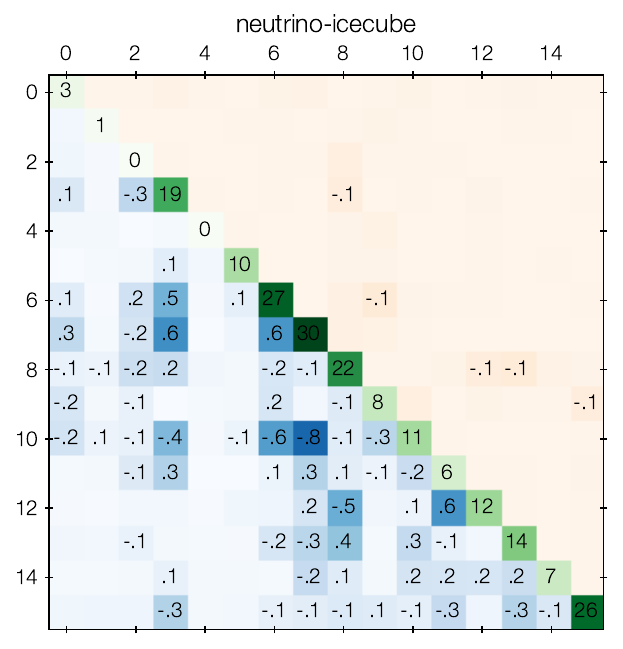}
\includegraphics[viewport=0bp 0bp 294bp 297bp,width=0.5\columnwidth]{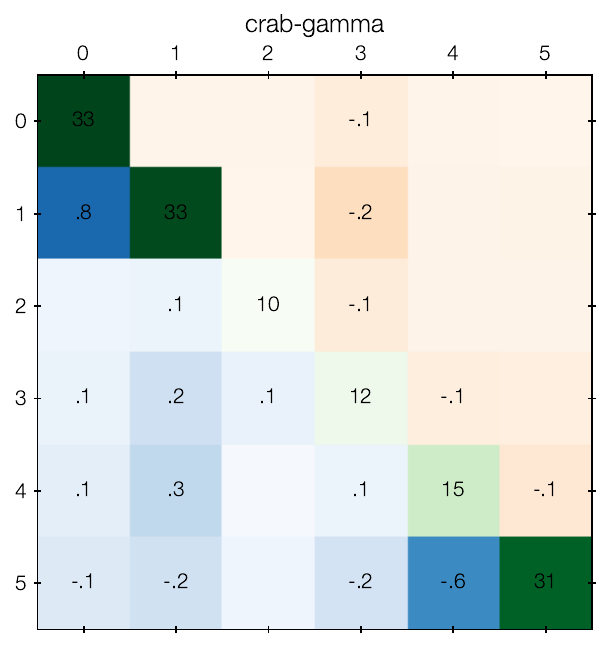}

\includegraphics[viewport=0bp 0bp 294bp 297bp,width=0.5\columnwidth]{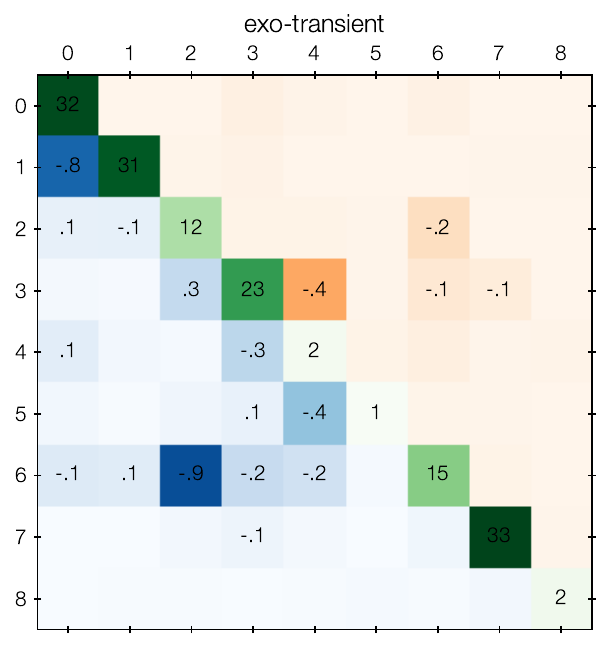}
\includegraphics[viewport=0bp 0bp 294bp 297bp,width=0.5\columnwidth]{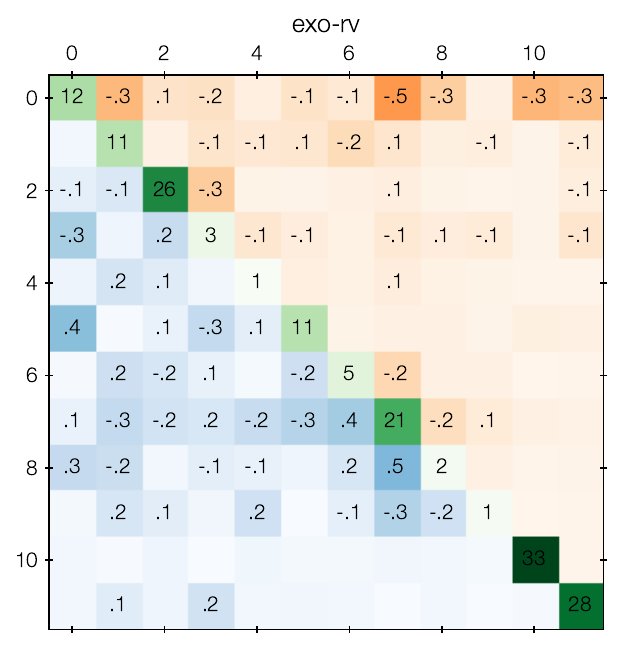}
\includegraphics[viewport=0bp 0bp 294bp 297bp,width=0.5\columnwidth]{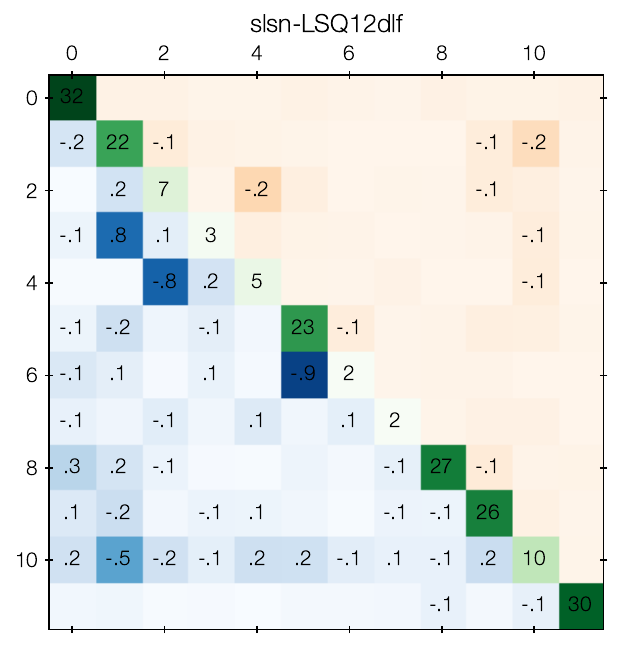}
\includegraphics[viewport=0bp 0bp 294bp 297bp,width=0.5\columnwidth]{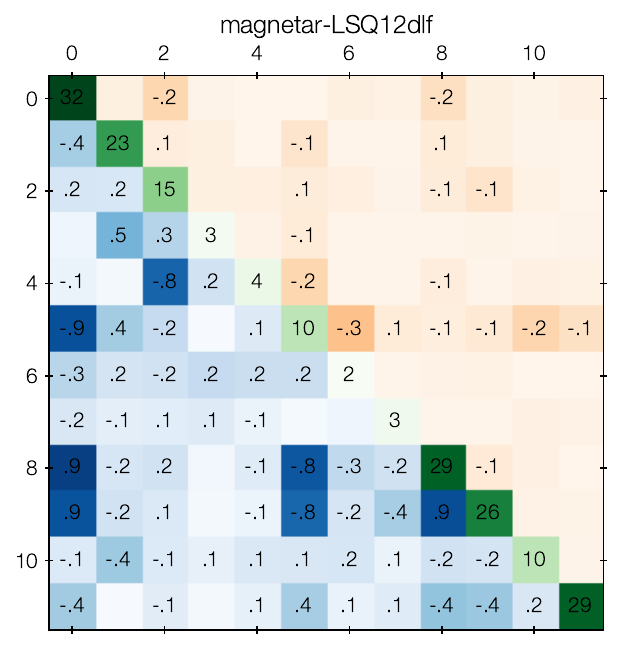}

\includegraphics[viewport=0bp 0bp 294bp 297bp,width=0.5\columnwidth]{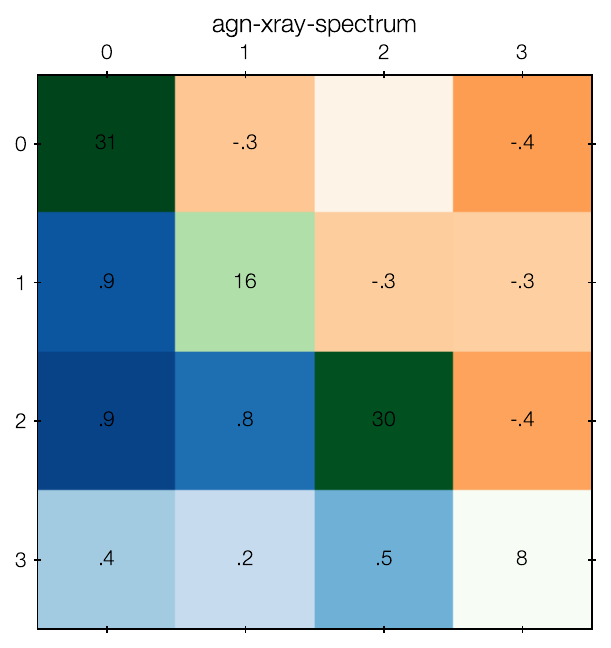}
\includegraphics[viewport=0bp 0bp 294bp 297bp,width=0.5\columnwidth]{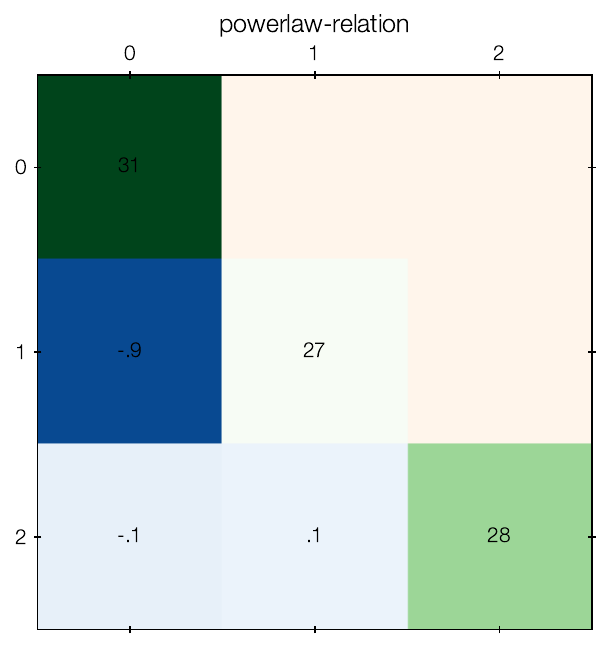}
\includegraphics[viewport=0bp 0bp 294bp 297bp,width=0.5\columnwidth]{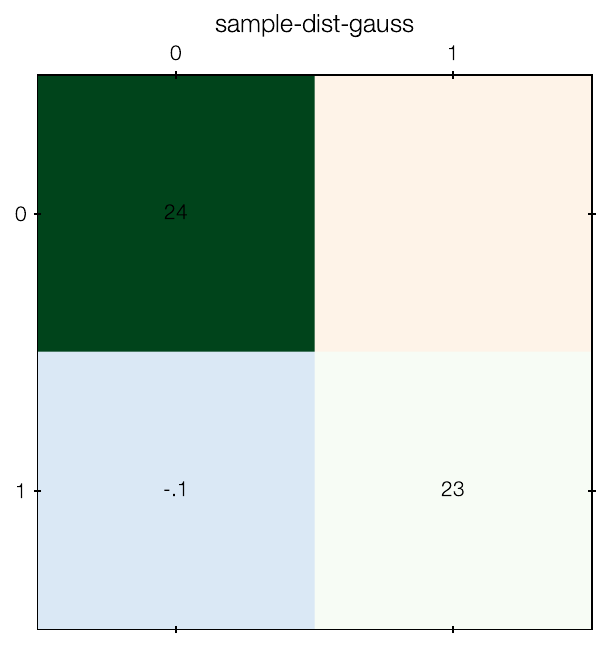}
\includegraphics[viewport=0bp 0bp 294bp 297bp,width=0.5\columnwidth]{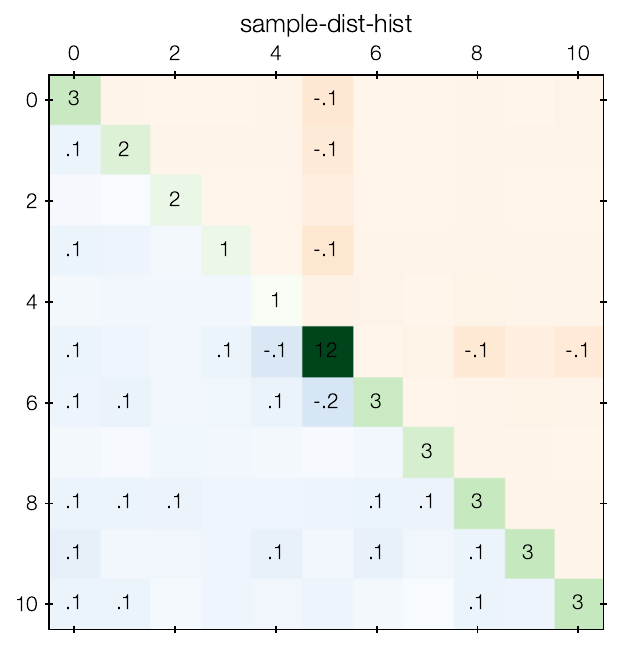}

\caption{\protect\label{fig:Posterior-structure-visualisatio}Posterior structure
visualisations (see section~\ref{subsec:Compact-visualisation-of})
for our real physics inference problems. The number of entries indicates
the dimensionality, diagonal entries indicate parameter information
gain, the bottom left triangle entries indicate linear parameter degeneracies
and the upper right triangle indicates non-linear degeneracies. For
example, the top left panel shows no non-linear degeneracy, while
the bottom left panel shows strong non-linear degeneracies.}

\end{figure*}

\begin{figure}

\includegraphics[viewport=0bp 0bp 391bp 442bp,width=1\columnwidth]{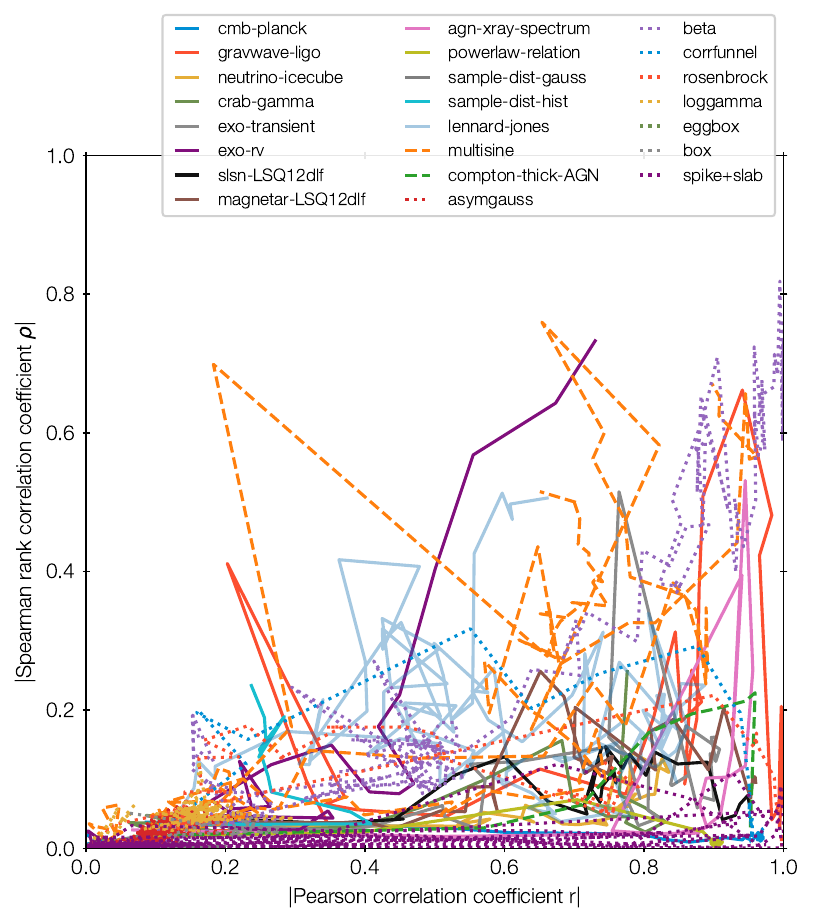}

\caption{\protect\label{fig:corr}For each problem, the evolution of the Pearson
correlation coefficient (x-axis) and the Spearman correlation coefficient
as nested sampling raises the likelihood threshold is shown as a curve.
Low coefficients (bottom left corner) indicate independent parameters,
high x-axis values indicate strong linear degeneracies. High y-axis
values indicate strong non-linear degeneracies, such as multiple modes,
or bananas in the parameter space.}
\end{figure}

The collated problems are listed in Table~\ref{tab:List-of-problems},
together with the derived characteristics. The model evaluation cost
is also listed, and ranges from less than 1ms to almost 1s per likelihood
evaluations. Figure~\ref{fig:Models} presents the location of each
inference problem in a corner plot. Among the ``real'' physics inference
problems (solid circles), there does not appear to a trend or preferred
region of the parameter space; all regions are covered. The mock problems
(crosses) cover the full parameter space as well.

The exoplanet radial velocity fitting with one planet shows a phase
transition, and heavy-tailed posteriors. These properties are also
produced by the spike and slab toy problem variations, calculated
here in only two dimensions. Fitting for 3 exoplanets yields 9 modes,
and the LSQ12dlf analyses show a similar number of modes. The multisine
mock problem mimics the properties of the exoplanet radial velocity
fitting in terms of multi-modality, width, parameter inequality, phase
transition and heavy-tailed posteriors, suggesting it is a useful
approximation.

As already shown in Figure~\ref{fig:volcurve}, the information gain
differs substantially. Additionally, the inequality column of Table~\ref{tab:List-of-problems}
indicates how much some parameters are learned while others are uninformative.
This parameter can be very high (>30).

Beyond those summary statistics, Figure~\ref{fig:Posterior-structure-visualisatio}
presents a visual impression of the posterior structure. Based on
the visualisation developed in section~\ref{subsec:Compact-visualisation-of},
this presents a diversity in dimensionality, as indicated by the number
of entries, the parameter information gain (diagonal entries), the
linear parameter degeneracies (bottom left triangles) and non-linear
degeneracies (upper right entries). Some inference problems, such
as cmb-planck (top left panel) show no non-linear degeneracy, while
others (agn-xray-spectrum, bottom left panel) show strong non-linear
degeneracies.

Finally, we have a deeper look at the evolving likelihood surface
as the sampler moves from the prior to the posterior mass. To this
end, we observe the live point distribution at regular snapshots.
We apply the same procedure as in Figure~\ref{subsec:Compact-visualisation-of}
and note the largest Pearson and (linearly whitened) Spearman correlation
coefficient. Figure~\ref{fig:corr} shows the evolution of these
values for each inference problem. For problems with no parameter
interactions, such as the asymgauss or beta, the curves remain in
the bottom left corner of the plot. Other problems however show strong
linear (Pearson) and/or non-linear (whitened Spearman) degeneracies
over the parameter space. This behaviour is stronger for real problems
(dashed and solid curves) than toy problems (dotted curves).

\section{Discussion \& future work}

\subsection{Strengths and limitations of the collected inference problems}

This work has assembled a list of parametric Bayesian inference applications
from astronomy, cosmology and particle physics. This is augmented
by mock applications and toy problems. A reproducible installation
is available at \footnote{\url{https://github.com/JohannesBuchner/space-of-inference-spaces/}},
allowing researchers to test with identical priors and likelihoods.
The parameter spaces and posterior distributions have been characterised
and cast into a space of parametric inference spaces.

This work focused on inference in astrophysics implemented in the
respective community-maintained packages. A cross-disciplinary effort
to compare the reliability of posterior distribution is posteriordb
\citep{Magnusson2021}, which contains 147 posteriors of 120 models
from diverse disciplines coded in Stan (and thus all differentiable).
In astrophysics, several works have compared the performance of Monte
Carlo inference algorithms \citep[e.g.,][]{Dunkley2005,Poppelaars2026,Albert2025}.
Typically, there is a strong focus on toy problems and only one inference
problem with real data from a particular sub-field. In this work,
we focused on understanding the nature of parameter spaces encountered
in inference across subfields of the space sciences.

A major result of this work is that the parameter space structure
of these inference spaces is highly diverse. This is illustrated in
Figure~\ref{fig:Models}. They span from low to high dimensionality,
can be mono-, bi- or multi-modal. A variety of complex posterior distribution
morphologies appear as illustrated in Figure~\ref{fig:Posterior-structure-visualisatio},
and inference on the parameters ranges from uninformative to extremely
informative. Furthermore, the computational cost of the physical models
can range from milliseconds to seconds. 

To test samplers in realistic, but controlled conditions with known
posteriors and evidence, toy inference problems are commonly used.
Standard ones include a Gaussian, Neal's funnel, log-gamma, eggbox,
rosenbrock and spike+slab, which are also included here. However,
toy inference problems may not represent the experience of a sampler
exploring the parameter space of a real data analysis problem. In
particular, Figure~\ref{fig:corr} demonstrates that real inference
problems can have stronger degeneracies than toy problems. There is
more work to be done to create toy inference problems that closely
mimic real physics inference. In this vein, we provide several small
data sets which are easy to work with yet represent real physics data
analysis (e.g., multisine, compton-thick-AGN, sample-dist, powerlaw-relation,
lennard-jones). In the future, perhaps the more complex data analysis
pipelines can be replaced by auxiliary functions, for example based
on Gaussian mixtures or normalising flows, which closely approximate
the real likelihood function, but have known, analytic properties
(integral Z, marginal distributions). Despite the limitations of artificial
toy problems, Figure~\ref{fig:Models} demonstrates that these cover
the same (vast) parameter space as the real inference problems (compare
the filled circles to crosses).

\subsection{Performance of samplers}

In recent years, a variety of Bayesian inference sampling packages
have been published. MCMC-based algorithms include emcee \citep{Foreman-Mackey2013},
Stan \citep{Carpenter2017}, zeus \citep{Karamanis2020} and pocoMC
\citep{Karamanis2022}. Nested sampling-based algorithms include multinest
\citep{Feroz2009}, polychord \citep{Handley2015a}, DNest4 \citep{Brewer2018DNest4},
dynesty \citep{Speagle2020}, ultranest \citep{UltraNest}, nautilus
\citep{Lange2023} and i-nessai \citep{Williams2023}. For reference
snowline\footnote{\url{https://johannesbuchner.github.io/snowline/}}
provides a Laplace's approximation implementation\footnote{based on \url{https://iminuit.readthedocs.io/}, see \citet{iminuit,James:1975dr}},
which can be improved with Gaussian mixtures learned by variational
Bayes and importance sampling \footnote{based on \url{https://github.com/pypmc/pypmc}, see \citet{Beaujean2013}}.

It is interesting to firstly try to understand in which regions of
the parameter space an inference algorithm gives reliable results.
This is achievable with problems where the truth is known (toy problems
or generated data sets). Only a subsequent step is performance comparison
among reliable algorithms. While the performance comparison is beyond
the scope of this work, we point out a few guidelines for comparing
Bayesian samplers fairly. Initially, the full sample of problems presented
here should be considered. If only a subset is of interest, this should
be clearly and transparently stated (e.g., focus on low-dimensional
inference problems, on mono-modal inference problems, etc.). It is
a trivial statement that with more compute budget, a better accuracy
and reliability can be achieved. Therefore, we recommend plotting
bias or accuracy against computational cost for each algorithm and
runtime. One quantifier of computational cost is the number of (likelihood)
model evaluations or wall-clock time. The former is suitable for computationally
challenging likelihoods, the latter is more relevant for computationally
cheap models paired with algorithms that require costly training (such
as deep neural networks).

It remains to be defined how to quantify the fidelity of the computation.
To quantify posterior fidelity, see for example the probability-probability
plots and Jensen-Shannon divergence quantification in \citet{Romero-Shaw2020}.
However, the reliable retrieval of $3\sigma$ upper or lower parameter
limits may also be of interest. Finally, when the true marginal likelihood
$Z$ is known and can be tested against, recovering $\ln Z$ to an
accuracy much better than $\sim0.3$ may not be useful in the context
of Bayes factors, as it only mildly affects the interpretation.

\subsection{Future work}

Breakthrough progress in machine learning is typically driven by (1)
open, large, high-quality data sets and (2) a clear formulation of
a meaningful objective. Examples span from MNIST\textquoteright s
challenge of digitizing hand-drawn numbers to the Critical Assessment
of Structure Prediction (CASP) \citep{Moult1995} protein-folding
challenge, recently met by AlphaFold \citep{Jumper2021}. The \textquotedblleft Learning
to learn by gradient descent by gradient descent\textquotedblright{}
paper \citep{Andrychowicz2016} demonstrated that optimization algorithms
can be derived by machine learning. Perhaps a similar breakthrough
can be accomplished for optimal Bayesian inference sampling procedures,
by providing an open, large data base and a clear objective. To this
end, the survey of inference problems encountered across cosmology,
particle physics, astrophysics and astronomy is presented. The representative
database includes fully specified likelihood and priors in a runnable
docker image with python interfaces. While similar previous work has
focused on providing simplified models that serve as unit tests \citep[e.g.,][]{inferencegym2020,Magnusson2021},
this work uses real-world inference with the software pipelines employed
by researchers.

As a first step, existing and novel algorithms can be judged across
the parameter space of problems for their empirical behaviour and
robustness, and to make well-founded recommendation for specific applications.
For more rapid testing and a unified interface, fast model emulators
for the provided real-world examples would be useful. This is left
for future work.

A step further into the future is comparable to Atari computer game
playing artificial intelligences \citep{Bellemare2012} that learn
optimal game playing strategies with reinforcement learning: A playground
for learning optimal Bayesian inference algorithms.
\begin{acknowledgements}
I thank Frederik Beaujean for insightful conversations, the referee,
Andrew Fowlie for constructive comments that improved the paper and
Jürgen Degenfellner for feedback on the reproducibility of the code.
I thank Gana Moharram, who studied the IceCube neutrino analysis for
a bachelor thesis co-supervised with Philipp Eller, for insightful
conversations.
\end{acknowledgements}

\bibliographystyle{aa}
\bibliography{agn,stats}

\appendix

\section{Real inference problems}

\label{subsec:Real-problems}

\subsection{Cosmology with the Cosmic Microwave Background}

(if you are a cosmologist, please help me with references!)

The Cosmic Microwave Background (CMB) is the oldest electromagnetic
signal observable. It originated when the Universe underwent a transition
from being so dense that photons would be constantly scattered to
the current state where photons can travel freely. These photons allow
us to measure the temperature of their regions of origin approximately
379,000 years after the Big Bang. A map of their emission over the
sky gives information about the temperature correlation. This carries
information of how the Big Bang inflation proceeded, which is dependent
on important constituents of the Universe, such as the baryonic matter
and dark matter content of the Universe. The CMB remains one of the
most important experiments to measure cosmological parameters in the
dark energy and cold dark matter cosmology framework ($\Lambda$CDM).
One difficulty in the fitting of cosmological models like $\Lambda$CDM
to the CMB is that the prediction of angular correlations is computationally
expensive (of the order of a few seconds per likelihood evaluation).

Here we adopt an example\cprotect\footnote{\url{https://github.com/JohannesBuchner/montepython_public/blob/3.5/input/example_ns.param}}
of the MontePython cosmology fitting package\cprotect\footnote{hosted at \url{https://github.com/brinckmann/montepython_public/}}.
We use the fake\_planck\_bluebook likelihood which emulates a Planck
measurement. The free parameters are $\Lambda$CDM cosmological parameters,
namely the baryonic density $\Omega_{\mathrm{b}}$ (between 1.8 and
3), the dark matter density $\Omega_{\mathrm{cdm}}$ (between 0.1
and 0.2), the scalar spectral index $n_{s}$ (between 0.9 and 1.1),
$A_{s}$ (between 1.8 and 3), the hubble parameter value, relative
to $100\,\mathrm{km/s/Mpc}$, $h$ (between 0.6 and 0.8), and the
time of reionisation $\tau_{\mathrm{reion}}$ (between 0.004 and 0.12).
All priors are uniform within the mentioned bounds.

\subsection{Gravitational wave analysis}

Gravitational waves originate from distortions of space-time by compact
objects. Recently, the development of multiple, extremely sensitive
instruments have allowed the observation of two black holes merging.
GW170817 \citep{Abbott2017} was the first gravitational event detected
by three detectors and allowed for the first time localisation on
the sky, albeit with substantial parameter degeneracies.

Here we adopt a tutorial example\cprotect\footnote{based on \url{https://github.com/gwastro/PyCBC-Tutorials/blob/7f5ff8fdd40c5dce2237b6082e1755b4aba9b989/tutorial/inference_1_ModelsAndPEByHand.ipynb}}
of the PyCBC gravitational wave analysis package \citep{Nitz2023pycbc}
. The merging system GW170817 is described by the mass ratio $q$,
the chirp mass $m_{\mathrm{chirp}}$, which is a combination of the
two masses that influences the signal amplitude, the inclination $i$
of the system relative to the observer, the time of coalescence $t_{c}$
in seconds, the distance $d$ and position on the sky ($RA$, $DEC$).
The priors adopted are:

\begin{align*}
q & \sim & \mathrm{Uniform}(1,2)\\
m_{\mathrm{chirp}} & \sim & \mathrm{Uniform}(1,2)\\
\sin i & \sim & \mathrm{Uniform}(0,1)\\
t_{c}-t_{m} & \sim & \mathrm{Uniform}(0.02,0.05)\\
d & \sim & \mathrm{Uniform}(10,100)\\
RA & \sim & \mathrm{Uniform}(0,2\pi)\\
\cos DEC & \sim & \mathrm{Uniform}(-1,1)
\end{align*}
where $t_{m}$ is the time automatically associated by an automated
pipeline searching for candidate events in the noise.

\subsection{Atmospheric Neutrino Oscillations from IceCube}

IceCube is a neutrino detector near the south pole, which has collected
atmospheric neutrino data \citep{Aartsen2018,Aartsen2019,Jia2019,Aartsen2020}
that can be studied for neutrino oscillation. Here we adopt a tutorial
example of the PISA IceCube analysis package\cprotect\footnote{based on \url{https://github.com/icecube/pisa/blob/f91224b58360ee9ecefe4bdb232249263c9eee17/pisa_examples/IceCube_3y_oscillations_example.ipynb}}
applied to three-year IceCube data, where the muon and neutrinos model
predicts a 2d-histogrammed signal, which is compared to collected
data. The parameters are adopted from the PISA defaults:

\begin{align*}
\mathrm{nue\_umu\_ratio} & \sim & \mathrm{Gauss}(1,\,0.05)\\
\mathrm{Barr\_uphor\_ratio} & \sim & \mathrm{Gauss}(0,\,1)\\
\mathrm{Barr\_nu\_nubar\_ratio} & \sim & \mathrm{Gauss}(0,\,1)\\
\mathrm{delta\_index} & \sim & \mathrm{Gauss}(0,\,0.1)\\
\mathrm{theta13} & \sim & \mathrm{Gauss}(8.5{^\circ},\,0.205{^\circ})\\
\mathrm{theta23} & \sim & \mathrm{Uniform}(31{^\circ},\,59{^\circ})\\
\mathrm{deltam31} & \sim & \mathrm{Uniform}(0.001\,\mathrm{eV},\,0.007\,\,\mathrm{eV})\\
\mathrm{aeff\_scale} & \sim & \mathrm{Uniform}(0,\,3)\\
\mathrm{nutau\_norm} & \sim & \mathrm{Uniform}(-1,\,8.5)\\
\mathrm{nu\_nc\_norm} & \sim & \mathrm{Gauss}(1,\,0.2)\\
\mathrm{opt\_eff\_overall} & \sim & \mathrm{Gauss}(1,\,0.1)\\
\mathrm{opt\_eff\_lateral} & \sim & \mathrm{Gauss}(25,\,10)\\
\mathrm{opt\_eff\_headon} & \sim & \mathrm{Uniform}(-5,\,2)\\
\mathrm{ice\_scattering} & \sim & \mathrm{Gauss}(0,\,10)\\
\mathrm{ice\_absorption} & \sim & \mathrm{Gauss}(0,\,10)\\
\mathrm{atm\_muon\_scale} & \sim & \mathrm{Uniform}(0,\,5)
\end{align*}

\subsection{Gamma-rays from the Crab pulsar-wind nebula}

The Crab nebula is the brightest astrophysical source in the sky at
high energies. The Large-Area Telescope (LAT) onboard Fermi has observed
emission from the region, which is a super-position of the Crab Pulsar
(PSR J0534+2200), and synchrotron and Inverse Compton emission from
the Crab Nebular. We follow the tutorial \cprotect\footnote{based on \url{https://threeml.readthedocs.io/en/stable/notebooks/Fermipy_LAT.html}}
of the 3ML multi-messenger package .

All parameters are assigned uninformative priors. The parameters include
the normalisation (uniform from 0 to $1.4\times10^{10}\mathrm{keV}^{-1}\mathrm{s}^{-1}\mathrm{cm}^{-2}$)
and spectral index (uniform from -10 to 10) of the ``super\_cutoff\_powerlaw''
model for PSR~J0534+2200, the normalisation for the power law spectrum
of NVSS~J052622+224801 (log-uniform from $10^{-20}$ to $1.1\times10^{-14}$)
and 4FGL~J0544.4+2238 (log-uniform from $10^{-20}$ to $1.39\times10^{-13}$),
and the isotropic background normalisation and the galactic background
factor. 

\subsection{Exoplanet transit observations}

Photometric light curves of stars can show small, periodic dips in
brightness, which are interpreted as transits of exoplanets in front
of the star. We follow the tutorial of the juliet package\cprotect\footnote{based on \url{https://juliet.readthedocs.io/en/latest/tutorials/transitfits.html#transit-fits}}
to analyse HATS-46 light curve observed with the Transiting Exoplanet
Survey Satellite \citep[see][for details]{Brahm2018}. The model is
essentially a constant light curve modified by a U-shaped dip that
repeats with some period. We use juliet \citep{Espinoza2019juliet}
for the implementation, which in turn uses batman \citep{Kreidberg2015}
for the light curve modelling together with limb-darkening \citep{Kipping2013}.

The model parameters include the properties of the planet (period
P in days, time-of-transit centre t0, planet-to-star radius ratio
p, eccentricity ecc, argument of periastron passage omega), as well
as nuisance parameters (dilution factor mdilution, offset relative
flux mflux, jitter sigma, Limb-darkening parameters q1 and q2):

\begin{align*}
\mathrm{P\_p1} & \sim & \mathrm{Gauss}(4.7,\,0.1)\\
\mathrm{t0\_p1} & \sim & \mathrm{Gauss}(1358.4,\,0.1)\\
\mathrm{r1\_p1} & \sim & \mathrm{Uniform}(0,\,1)\\
\mathrm{r2\_p1} & \sim & \mathrm{Uniform}(0,\,1)\\
\mathrm{q1\_TESS} & \sim & \mathrm{Uniform}(0,\,1)\\
\mathrm{q2\_TESS} & \sim & \mathrm{Uniform}(0,\,1)\\
\mathrm{ecc\_p1} & = & 0\\
\mathrm{omega\_p1} & = & 90\\
\mathrm{rho} & \sim & \mathrm{LogUniform}(100,\,10000)\\
\mathrm{mdilution\_TESS} & = & 1.0\\
\mathrm{mflux\_TESS} & \sim & \mathrm{Gauss}(0,\,0.1)\\
\mathrm{sigma\_w\_TESS} & \sim & \mathrm{LogUniform}(0.1,\,1000)
\end{align*}

\subsection{Exoplanet detection from Radial Velocity data}

\begin{figure*}

\includegraphics[width=1\columnwidth]{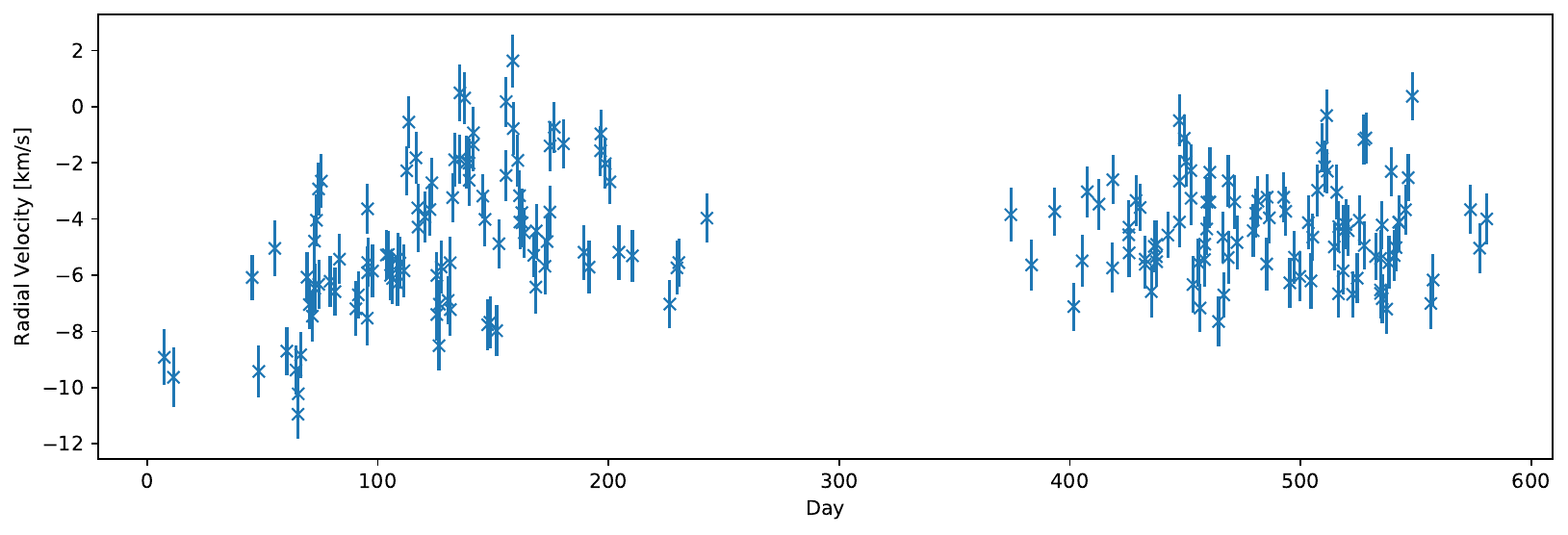}

\caption{\protect\label{fig:Sine-data-1}Exoplanet Doppler shift time series
data (blue).}
\end{figure*}

One of the most efficient ways to detect exoplanets is through changes
in the line-of-sight (radial) velocity of individual stars, as the
tug of planets gravitationally accelerates them (Doppler shift). A
planetary system leads to a complex overlay of periodic velocity changes.
Additionally, the measurement of velocities is uncertain, in part
because of the instrument accuracy and precision, and in part because
the spectral emission lines used to measure Doppler shifts can be
unstable due to stellar activity. That latter process has been modeled
with Gaussian processes in recent years.

Here we adopt the problem setup of the Extreme Precision Radial Velocity
III challenge\footnote{hosted at \url{https://github.com/EPRV3EvidenceChallenge/Inputs}}
\citep{Nelson2020}. They simulated several artificial exoplanet systems
containing two planets, with a Gaussian process and realistic observation
time sampling. The challenge participants were asked to compute the
Bayesian marginal likelihoods of each data set assuming that the true
number of planets was 0, 1, 2 or 3. Participants did not know the
true simulation input, but they were told the exact Gaussian noise
properties, which follows a Gaussian process. Here we repeat this
exercise, for their dataset 5 (shown in Figure~\ref{fig:Sine-data-1}).
The exact specification of this problem is defined in \citep{Nelson2020}.
Essentially, it is similar to a sine time series fit, except that
the periodic signal can be asymmetric due to ellipticity, giving 5
parameters per planet (signal amplitude, period, pericentre time,
eccentricity and mean anomaly) that describe a Keplerian orbit. Additionally,
the white noise amplitude $\sigma_{j}$ and the systematic velocity
$C$ are free parameters, giving $2+N_{\mathrm{planets}}\times5$
free parameters for $N_{\mathrm{planets}}=0,1,2,3$. We use juliet
\citep{Espinoza2019juliet} for the implementation, which in turn
uses RadVel \citep{Fulton2018} for Kepler orbits and george \citep{Ambikasaran2014}
for modelling Gaussian processes.

\subsection{X-ray spectral analysis of Active Galactic Nuclei}

Active Galactic Nuclei (AGN) are regions in the centres of massive
galaxies where super-massive black holes grow. As gas swirls into
the black hole, enormous amounts of radiation are produced by release
of gravitational energy, sometimes shining brighter than all host
galaxy stars together. Close to the black hole, X-rays are also produced
and are an important tracer of the mass inflow into the black hole.
They also allow identifying AGN in the sky, even when the black hole
is surrounded by thick gas and dust, as most of the energetic X-rays
penetrate through any obscurers. Space-based X-ray focusing instruments
allow measurements of the X-ray spectra, which carry information on
the AGN luminosity, obscuring column density and properties of the
X-ray emitter (photon index, energy turn-over). The detection of X-ray
radiation is performed by counting photon events and capturing an
estimate of their energy, time of arrival and location on the sky.
However, the energy response and energy-dependent sensitivity of the
instrument adds some analysis complexity. Additionally, when few counts
are detected per energy bin, as is commonly the case, the process
needs to be modelled with a Poisson likelihood. For more details,
see \citet{Buchner2014} and \citet{Dyk2001}.

Here, we include the source of a Chandra deep observation (source
ID 179) presented \citet{Buchner2014}. The data are available online\footnote{\url{https://github.com/JohannesBuchner/BXA/tree/master/examples/sherpa}}
and the model is defined by the xagnfitter script\footnote{\url{https://johannesbuchner.github.io/BXA/xagnfitter.html}}
of the Bayesian X-ray analysis package \citep[BXA;][]{Buchner2021}
based on sherpa \citep{Freeman2001}. The intrinsic X-ray emission
is a power law, which is modulated by a physical obscurer model simulation
of photo-electric absorption, Compton-scattering and fluorescent line-emission
\citep{Buchner2019a}. This model is available as a table\footnote{\url{https://github.com/JohannesBuchner/xars/blob/master/doc/uxclumpy.rst}}.
To this obscured AGN model, a soft, unobscured power law is added.
The background spectrum is added and its normalisation determined
simultaneously with a joint fit of the spectrum extracted in a background
sky region and the source+background spectrum extracted at the source
location. The model has four parameters. This includes a wide log-uniform
prior (between $10^{-8}$ and $10^{-3}$) on the power law normalisation,
a wide log-uniform prior (between $10^{-7}$ and $10^{-1}$) on the
soft power law normalisation relative to the primary power law normalisation,
intrinsic power law, a Gaussian prior on the power law photon index
with mean 1.95 and standard deviation 0.15, a log-uniform prior on
the obscurer column density between $10^{20}$ and $10^{26}/\mathrm{cm}^{2}$.

\subsection{Superluminous supernova}

Super-Luminous Supernovae (SLSN) are extreme explosions at the end
of stellar evolution. By fitting photometric observations over time,
the explosion mechanism and its physical parameters can be determined
(and distinguished from other transients). We follow the tutorial
of MOSFiT \cprotect\footnote{based on \url{https://mosfit.readthedocs.io/en/latest/fitting.html#public-data}},
to analyse LSQ12dlf, with two models described in \citet{Nicholl2017}:
a magnetar engine with a simple spectral energy distribution (`magnetar`
model) and a magnetar with a modified spectrum and additional constraints
(`slsn` model). The parameters are:

\begin{align*}
\mathrm{nhhost} & \sim & \mathrm{LogUniform}(10^{16},\,10^{23})\\
\mathrm{Pspin} & \sim & \mathrm{Uniform}(1,\,10)\\
\mathrm{Bfield} & \sim & \mathrm{LogUniform}(0.1,\,10)\\
\mathrm{Mns} & \sim & \mathrm{Uniform}(1,\,2)\\
\mathrm{thetaPB} & \sim & \mathrm{Uniform}(0,\,1.5708)\\
\mathrm{texplosion} & \sim & \mathrm{Uniform}(-500,\,0)\\
\mathrm{kappa} & \sim & \mathrm{Uniform}(0.05,\,0.2)\\
\mathrm{kappagamma} & \sim & \mathrm{LogUniform}(0.1,\,10000)\\
\mathrm{mejecta} & \sim & \mathrm{LogUniform}(0.001,\,100)\\
\mathrm{vejecta} & \sim & \mathrm{Uniform}(5000,\,20000)\\
\mathrm{temperature} & \sim & \mathrm{LogUniform}(1000,\,100000)\\
\mathrm{variance} & \sim & \mathrm{LogUniform}(0.001,\,100)\\
\mathrm{codeltatime} & \sim & \mathrm{LogUniform}(0.001,\,100)\\
\mathrm{codeltalambda} & \sim & \mathrm{LogUniform}(0.1,\,10000)
\end{align*}
In case of `slsn`, the following parameters differ:
\begin{align*}
\mathrm{Bfield} & \sim & \mathrm{Uniform}(0.1,\,10)\\
\mathrm{mejecta} & \sim & \mathrm{LogUniform}(0.1,\,100)\\
\mathrm{vejecta} & \sim & \mathrm{Uniform}(5000,\,20000)\\
\mathrm{temperature} & \sim & \mathrm{Uniform}(3000,\,10000)
\end{align*}

\subsection{Lennard-Jones potential}

In material science, the group behaviour of atoms gives rise to structures
and phases of matter (solids, liquids, gases, crystals, metals, etc).
A standard example is the Lennard-Jones potential, which we adopt
here with 6 particles in a box. Pairs of particles feel two forces,
a repulsive force at short distances (Pauli repulsion) and an attractive
force at longer distances (van der Waals force). This are formulated
as:

\[
{\cal L}=\prod_{i}\prod_{j>i}\left(\frac{\sigma}{r_{ij}}\right)^{6}-\left(\frac{\sigma}{r_{ij}}\right)^{12}
\]
where $r_{ij}$ is the Euclidean distance between particles with indices
$i$ and $j$, but at least $\sigma$. We set $\sigma=10^{-3}$, which
defines the scale at which particles feel attraction or repulsion. 

Each particle's three-dimensional location (x,y,z) is a free parameter
between -1 and +1. To avoid identical modes, we set the likelihood
to zero when $|z_{1}|<|z_{2}|<...<|z_{n}|$ is not satisfied. Due
to translation symmetry, we assume the first particle is placed in
the coordinate centre (0,0,0). From rotation symmetry, we assume the
second particle is placed along the positive z direction (0,0,$z_{2}$).
From a second rotation symmetry, the third particle is placed on the
(0,$y_{3}$,$z_{3}$) plane. The remaining particles have all (x,y,z)
coordinates as free parameters.

\subsection{Power-law line fit: Tully-Fisher relation}

Stars in the centres of galaxies move akin to a ideal gas, so that
the velocity dispersion of galactic central elliptical components
(bulges) correlates with the bulge mass. \citet{Kormendy2013} presented
a compilation of measurements of velocity dispersions and masses,
both annotated with asymmetric (and heteroscedastic) error bars. Following
the ``Fitting a line'' tutorial\footnote{\url{https://johannesbuchner.github.io/UltraNest/example-line.html}}
of UltraNest, we assume these correspond to Gaussian tails, scaled
according to the error bar size. The measurements are fit with a power
law, with the intrinsic scatter along the power law accounted for
by a log-Normal distribution. The parameters are

\begin{align*}
\mathrm{slope} & \sim & \mathrm{Uniform}(-3,\,3)\\
\mathrm{offset} & \sim & \mathrm{LogUniform}(10,\,1000)\\
\mathrm{scatter} & \sim & \mathrm{LogUniform}(0.001,\,10)
\end{align*}
The likelihood is integrating the Log-Normal distribution over the
data point distribution, i.e., it is a hierarchical Bayesian model.
This integration is performed numerically.

\subsection{Sample distributions: A galaxy without dark matter}

\subsubsection{Gaussian Sample distribution}

\citet{vanDokkum2018} observed the velocity of globular clusters
in a low-mass, ultra-diffuse galaxy. The width of the (Gaussian) distribution
of velocities, i.e., the velocity dispersion, is directly related
to the total galaxy mass. From comparing the stellar light to the
total light, they inferred that the dark matter mass in this galaxy
is negligible. Following \footnote{\url{https://johannesbuchner.github.io/UltraNest/example-intrinsic-distribution.html}},
we analyse their velocity measurements with Gaussian error bars, assuming
a Gaussian sample distribution (see e.g., \citealp{Baronchelli2018}
for such a Gaussian hierarchical model in astronomy). The parameters
are the mean, for which we assume a uniform distribution between -100
and 100 km/s, and the scatter, for which we assume a log-uniform distribution
between 1 and 1000 km/s. The parameter for each data point's true
value is integrated out, as implemented in \footnote{\url{https://github.com/JohannesBuchner/PosteriorStacker}}.

\subsubsection{Dirichlet sample distribution}

The above example is repeated, but with a more flexible sample distribution.
Following the PosteriorStacker tutorial\footnote{\url{https://github.com/JohannesBuchner/PosteriorStacker}},
a uniform Dirichlet distribution is adopted, with 11 uniformly spaced
bins between -80 km/s and +80 km/s. This analysis allows checking
whether a Gaussian was a reasonable model.

\section{Mock problems\protect\label{sec:Mock-problems}}

\subsection{Sine time series}

\begin{figure}

\includegraphics[width=1\columnwidth]{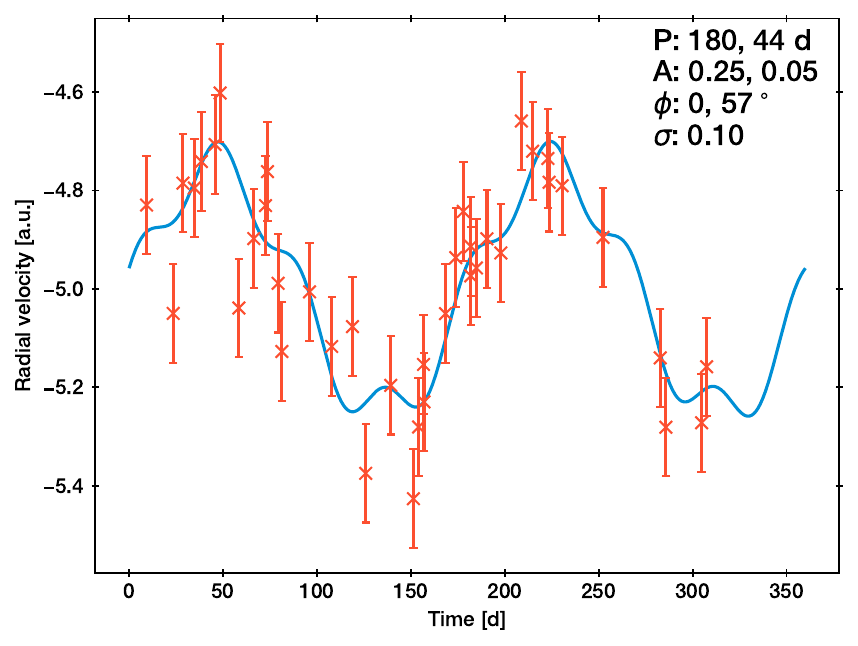}

\caption{\protect\label{fig:Sine-data}Sine time series data (orange) with
generating two-component model.}
\end{figure}

Analyses of inhomogeneously sampled light curves is a common problem
in astrophysics, sometimes with complex noise processes and semi-periodic
signals. Here we present a simple multi-component sine signal as a
toy problem, and uniformly randomly sample observing times. The problem
is defined as:

\begin{align*}
\log L & = & \prod_{i=1}^{M}\mathrm{Normal}(m(t_{i},\theta)-d_{i},\sigma^{2})\\
m(t,\theta) & = & m_{0}+\sum_{j=1}^{n_{\mathrm{comp}}}A_{c}\times\sin\left(\frac{2\pi t}{P_{j}}+\phi_{j}\right)\\
m_{0} & \sim & \mathrm{Uniform}(-50,50)\\
\log\sigma & \sim & \mathrm{Uniform}(-1.5,-0.5)\\
\log A & \sim & \mathrm{Uniform}(-2,2)\\
\phi & \sim & \mathrm{Uniform}(0,2\pi)\\
\log P & \sim & \mathrm{Uniform}(-1,3)
\end{align*}
Figure~\ref{fig:Sine-data} presents a graph $(t_{i},d_{i})$ of
$M=40$ data points, generated by sampling $n_{\mathrm{comp}}=2$
components with offset $m_{0}=-5$, measurement uncertainty $\sigma=0.1$,
amplitudes $A_{1}=0.25$, $A_{2}=0.05$, periods $P_{1}=180$, $P_{2}=44$
and phases $\phi_{1}=0$, $\phi_{2}=1\,\mathrm{rad}$. The same data
are analysed with $n_{\mathrm{comp}}=0,\,1,\,2,\,3$, producing two
to eleven dimensional problems with completely exchangable components.

\subsection{Low X-ray count observations of a Compton-thick AGN}

\begin{figure*}

\includegraphics[width=1\columnwidth]{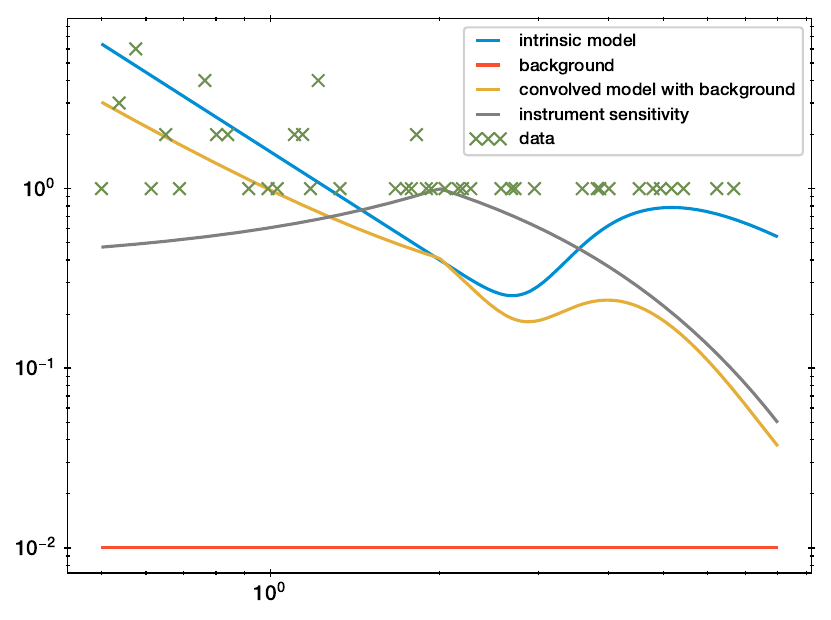}

\caption{\protect\label{fig:xraydata}X-ray spectrum with the true generating
model. The total model (blue curve) is a sum of two power laws, one
altered by a exponential truncation towards the low-energy side. That
model is multiplied by the instrument sensitivity (gray curve), and
background contamination (red) is added. The final model is shown
in blue, from which poisson data are drawn (green crosses).}
\end{figure*}

This mock problem (`xrayspectrum`) describes a 5-dimensional degenerate
physical parameter space discussed in \citet{Buchner2014} of a heavily
obscured active galactic nucleus. The model components and data are
described in Figure~\ref{fig:xraydata}. The emission spectrum over
200 energy channels from 0.5 to 8~keV is described by:
\[
F(E)=B+A_{LE}(E)\times E^{-\Gamma}\times\left[\exp\left(-N_{\mathrm{H}}\times E^{-3}\right)+f_{\mathrm{scat}}\right]
\]
with background amplitude $B$, signal amplitude $A$, spectral photon
index $\Gamma$, obscuring column density $N_{H}$, and fraction of
the power law that escapes unobscured $f_{\mathrm{scat}}$. The instrument
sensitivity peaks near 2keV: $A_{LE}(E)=\exp\left\{ -\left|\frac{E-2\mathrm{keV}}{2\mathrm{keV}}\right|\right\} $.

This model shows a degeneracy when $A$ is low and comparable to $B$,
because high $N_{H}$ and high $f_{\mathrm{scat}}$ look like $N_{H}=0$
(a simple power law). The model is illustrated in Figure~\ref{fig:xraydata}.
The priors on the parameters are:

\begin{align*}
\log A & \sim & \mathrm{Uniform}(-5,+5)\\
\Gamma & \sim & \mathrm{Normal}(2,0.2^{2})\\
\text{\ensuremath{\log N_{\mathrm{H}}}} & \sim & \mathrm{Uniform}(-3,+3)\\
\log f_{\mathrm{scat}} & \sim & \mathrm{Uniform}(-7,-1)\\
\ln B & \sim & \mathrm{Normal}(0.2,0.1^{2})
\end{align*}

\section{Toy problems\protect\label{sec:Toy-problems}}

\label{subsec:Toy-problems}

In this section, the prior is the unit hypercube ($0<\theta_{i}<1$
for $1\leq i\leq d$), unless specified otherwise. If analytically
known, the marginal parameter posterior distribution $p(\theta_{i}|D)$
and marginal likelihood $Z$ are given.

\subsection{Asymmetric Gaussian}

Integration of a Gaussian distribution is a standard test problem.
The variation here introduces some parameter inequality and spreads
the means in a sine pattern, so that the posterior is not at the centre
of the prior range.

\begin{align*}
L & = & \prod_{i=1}^{d}\mathrm{Normal}(\mu_{i},\sigma_{i}^{2})\\
\sigma_{i} & = & 0.1\times10^{-\left(-9-\frac{\sqrt{d}}{2}\right)\times\frac{i-1}{d-1}}\\
\mu_{i} & = & \frac{1}{2}+\frac{1-5\sigma_{i}}{2}\times\sin\frac{i-1}{2d}
\end{align*}
This problem is evaluated using uniform priors on $d=4$ (making $\sigma_{i}$
range from $10^{-9}$ to $10^{-1}$), $16$ ($10^{-8}<\sigma_{i}<10^{-1}$)
and $100$ dimensions ($10^{-5}<\sigma_{i}<10^{-1}$). The four-dimensional
case is shown in the top left panel of Figure~\ref{fig:pairwise-posterior}.
Note the different axes ranges. The true posterior is $p(\theta_{i}|D)=\mathrm{Normal}(\mu_{i},\sigma_{i}^{2})$,
and the marginal likelihood $Z\approx1$.

\subsection{Beta product}

Diverse test problems can be generated by combining standard one-dimensional
probability distributions. Here, the Beta distribution is used to
represent diverse posterior shapes in each parameter:

\begin{align*}
L & = & \prod_{i=1}^{d}\mathrm{Beta}(a_{i},b_{i})
\end{align*}
with fixed, known $a$,$b$, randomly generated as:

\begin{align*}
\log a_{i} & \sim & \mathrm{Uniform}(-1,1)\\
\log b_{i} & \sim & \mathrm{Uniform}(-1,1)
\end{align*}
This distribution can produce multiple modes (where $a_{i}<1$ and
$b_{i}<1$), non-Gaussian tails. The likelihood is relatively uninformative
on each parameter (completely non-informative when $a_{i}=b_{i}=0$).
We test this problem in 2, 10 and 30 dimensions. The true marginal
posteriors are given by $P(\theta_{i}|D)=\mathrm{Beta}(a_{i},b_{i})$,
and the marginal likelihood is $Z=1$.

\subsection{Correlated Funnel}

Neil's funnel is a standard test problem that represents features
of hierarchical Bayesian models. It is a normal distribution with
the standard deviation also a free parameter. This causes a funnel
shape (see middle panel of Figure~\ref{fig:pairwise-posterior})
involving all parameters. Such non-affine correlations can sometimes
be eased significantly by reparametrizations which scale the unknown
mean parameters by the standard deviation parameter \citet{Betancourt2013}.
Here we use the correlated version of \citet{Karamanis2020}:

\begin{align*}
L & = & \prod_{i}\mathrm{Normal}(\mu_{i}-\gamma\times\mu_{i-1},\Sigma^{2})\\
\ln\sigma & \sim & \mathrm{Normal}(0,1)\\
\mu_{i} & \sim & \mathrm{Uniform}(-100,100)\\
\Sigma_{ij} & = & \begin{cases}
\sigma & \mathrm{if\,}i=j\\
\gamma\times\sigma & \mathrm{otherwise}
\end{cases}
\end{align*}
This problem is tested in 2, 10 and 50 dimensions with correlation
strength $\gamma=0.95$. The true marginal posteriors are $p(\mu_{i}|D)=\mathrm{Normal}(0,1)$,
$p(\ln\mu_{i}|D)=\mathrm{Normal}(0,1)$, and the marginal likelihood
$Z\approx1$.

\subsection{Rosenbrock function}

The Rosenbrock function is a standard test problem in optimization.
It exhibits a non-linear, narrow degeneracy that can be difficult
to navigate (right middle panel of Figure~\ref{fig:pairwise-posterior}).
We adopt a probabilistic formulation suggested by \citet{RosenbrockChallenge}
(see a similar version in \citet{Jia2019}):

\begin{align*}
\log L & = & -2\times\sum_{i=1}^{d-1}100\times\left(\theta_{i+1}-x_{i}^{2}\right)^{2}+\left(1-\theta_{i}\right)^{2}\\
\theta_{i} & \sim & \mathrm{Uniform}(-10,10)
\end{align*}
We test this problem in 2, 20 and 50 dimensions.

\subsection{LogGamma}

The LogGamma problem \citet{Beaujean2013} exhibits multi-modality
and heavy tails, which lead to non-elliptical contours.

\begin{eqnarray*}
g_{a} & \sim & \mathrm{LogGamma}\left(1,\,\frac{1}{3},\,\frac{1}{30}\right)\\
g_{b} & \sim & \mathrm{LogGamma}\left(1,\,\frac{2}{3},\,\frac{1}{30}\right)\\
n_{c} & \sim & \mathrm{Normal}\left(\frac{1}{3},\,\frac{1}{30}\right)\\
n_{d} & \sim & \mathrm{Normal}\left(\frac{2}{3},\,\frac{1}{30}\right)\\
d_{i} & \sim & \mathrm{LogGamma}\left(1,\,\frac{2}{3},\,\frac{1}{30}\right)\,\text{\,\,\ if\,\,\,}3\leq i\leq\frac{d+2}{2}\\
d_{i} & \sim & \mathrm{Normal}\left(\frac{2}{3},\,\frac{1}{30}\right)\,\text{\,\,\,\,\,\,\,\,\,\,\,\,\,\,\,\,\,\,\,\,\ if\,\,\,}\frac{d+2}{2}<i\\
L_{1} & = & \frac{1}{2}\left(g_{a}(x_{1})+g_{b}(x_{1})\right)\\
L_{2} & = & \frac{1}{2}\left(n_{c}(x_{2})+n_{d}(x_{2})\right)\\
L & = & L_{1}\times L_{2}\times\prod_{i=3}^{d}d_{i}(x_{i})
\end{eqnarray*}
We test this problem in 2, 10 and 30 dimensions. The true marginal
posteriors are given by $P(x_{1}|D)=L_{1}(x_{1})$, and the marginal
likelihood is $Z=1$.

\subsection{Eggbox}

The eggbox function is a two-dimensional extremely multi-modal function
(bottom left panel of Figure~\ref{fig:pairwise-posterior}) proposed
by \citet{Feroz2008}, defined as: 

\begin{align*}
\log L & = & \left(2+\cos(5\pi\cdot\theta_{1})\cdot\cos(5\pi\cdot\theta_{2})\right)^{5}\\
\theta_{i} & \sim & \mathrm{Uniform}(0,10\pi)
\end{align*}

\subsection{Box}

This is a flat distribution at the corner of the parameter space,
placed on top of a wide, unimportant Gaussian distribution:

\begin{align*}
\ln L & = & -\frac{1}{2}\times\left(\frac{\theta}{0.1}\right)^{2}+100\times I[\delta<0.1]
\end{align*}
Here, $\delta=\max_{i}|\theta_{i}|$ is the parameter with the largest
deviation from zero. The priors are the standard uniform distribution
on all parameters $\theta_{i}$.

We test this problem in $d=5$ dimensions. The true marginal posteriors
are given by $P(\theta_{i}|D)=\mathrm{Uniform}(0,0.1)$, and the marginal
likelihood is $Z\approx100+(0.1)^{d}$. 

\subsection{Spike and slab}

The spike and slab problem is a mixture of two Gaussians, one with
a wide standard deviation, one with a narrow standard deviation. This
leads to a strong phase transition. The narrower Gaussian has standard
deviation $\sigma_{2}=f^{-\frac{1}{d}}$, and is shifted by $\Delta$
in each direction. The wider Gaussian has standard deviation $\sigma_{1}=1$.
Then the likelihood is:
\begin{eqnarray*}
L & = & L_{1}+L_{2}\\
L_{1} & = & \frac{w_{1}}{1+w_{1}}\prod_{i}\frac{1}{2\pi\sigma_{1}^{2}}\exp\left\{ -\frac{1}{2}\times\left(\frac{\theta_{i}-\Delta\times\sigma_{1}}{\sigma_{1}}\right)^{2}\right\} \\
L_{2} & = & \frac{1}{1+w_{1}}\prod_{i}\frac{1}{2\pi\sigma_{2}^{2}}\exp\left\{ -\frac{1}{2}\times\left(\frac{\theta_{i}}{\sigma_{2}}\right)^{2}\right\} 
\end{eqnarray*}
We use a two-dimensional setup, with the two Gaussians co-located
$\Delta=0$. For the relative weights we adopt $w_{1}$ values of
1, 40, 1000, and for $\sigma_{2}$ adopt 4, 40, 400 or 4000. This
gives 16 mono-modal toy problems with phase transitions. Secondly,
we adopt offset values for $\Delta$ of 1, 2, 4, 10 with weights $w_{1}$
of 1, 40 or 1000. This gives another 8 (bimodal) toy problems. The
problems are named ``spikeslab-$w_{1}$-$d$d-$\sigma_{2}$(-off$\Delta$)''.
The parameters are uniformly distributed from -10 to +10.

The true evidence is $Z=20^{-d}\approx e^{-6}$ when $\sigma_{2}$
is small, and the true marginals are the superposition of the two
weighted Gaussians. 
\end{document}